\documentclass[conference]{IEEEtran}
\ifCLASSINFOpdf
\else
\fi

\usepackage{comment}
\usepackage{makeidx}  
\usepackage{graphicx}
\usepackage[usenames, dvipsnames]{color}
\usepackage[]{algorithm2e}
\usepackage[justification=centering]{caption}
\usepackage{amsmath}
\usepackage{multirow}
\usepackage{multicol}
\usepackage{subcaption}

\begin{document}
%
\title{Killing Two Birds with One Stone: Malicious Domain Detection with High Accuracy and Coverage}

\author{\IEEEauthorblockN{Issa Khalil, Bei Guan, Mohamed Nabeel, Ting Yu}
\IEEEauthorblockA{Qatar Computing Research Institute\\
\{ikhalil,bguan,mnabeel,tyu\}@hbku.edu.qa}
}


%


\maketitle

\begin{abstract}
Inference based techniques are one of the major approaches to analyze DNS data and detecting malicious domains. The key idea of inference techniques is to first define associations between domains based on features extracted from DNS data. Then, an inference algorithm is deployed to infer potential malicious domains based on their direct/indirect associations with known malicious ones. The way associations are defined is key to the effectiveness of an inference technique. It is desirable to be both accurate (i.e., avoid falsely associating domains with no meaningful connections) and with good coverage (i.e., identify all associations between domains with meaningful connections). Due to the limited scope of information provided by DNS data, it becomes a challenge to design an association scheme that achieves both high accuracy and good coverage. 
%

In this paper, we propose a new association scheme to identify domains controlled by the same entity. Our key idea is an in-depth analysis of active DNS data to accurately separate public IPs from dedicated ones, which enables us to build high-quality associations between domains. Our scheme avoids the pitfall of naive approaches that rely on weak ``co-IP'' relationship of domains (i.e., two domains are resolved to the same IP)  that results in low detection accuracy, and, meanwhile, identifies many meaningful connections between domains that are discarded by existing state-of-the-art approaches. Our experimental results show that the proposed association scheme not only significantly improves the domain coverage compared to existing approaches but also achieves better detection accuracy.

Existing path-based inference algorithm is specifically designed for DNS data analysis. It is effective but computationally expensive. To further demonstrate the strength of our domain association scheme as well as improving inference efficiency, we investigate the effectiveness of combining our association scheme with the generic belief propagation algorithm. Through comprehensive experiments, we show that this approach offers significant efficiency and scalability improvement with only minor negative impact of detection accuracy, which suggests that such a combination could offer a good tradeoff for malicious domain detection in practice.
\end{abstract}


%

\section{Introduction}\label{introduction}
DNS data is one of the most notable sources of information utilized to detect malicious domains~\cite{PassiveDnsReplication_Weimer2005,Thales_Kountouras2016}. In general, there are two types of approaches that complement each other. In classification-based approaches, a classifier is built from local features of domains extracted from DNS data, which may be further enriched with other network and host features. A classifier is then trained using a ground truth dataset of benign and malicious domains, and used to classify new unknown domains. Inference-based approaches, which is the focus of this paper, are centered on building associations between domains from DNS data to reflect their meaningful connections (e.g., deployed and controlled by the same entity). Once such associations are defined, an inference algorithm is deployed to reason the maliciousness of a domain based on its direct/indirect associations with known malicious ones.

Inference-based approaches are based on a simple intuition: if a domain has strong associations with known malicious domains, it is likely to be malicious as well. Clearly how such associations are defined is key to the effectiveness of malicious domain detection. Ideally, it should satisfy two properties. First, it should be {\em accurate}, in the sense that the connections identified are indeed strongly relevant to maliciousness inferences. For example, it is reasonable to define associations when two domains are controlled and deployed by the same entity: if one domain is malicious, it is probably either directly deployed by or compromised by an attacker. In either case, it raises the probability of maliciousness of the other domain. As a contrast, it would not be reasonable to associate two domains just because they start with the same character. Associations based on irrelevant or weak connections would result in high false positives in malicious domain detection. Second, associations should be defined to have good {\em coverage}:  
it should reveal as many relevant connections between domains as possible. An overly strict association (e.g., two domains must have the same owner in their WHOIS records and the same authoritative server, be with the same subdomain names, and meanwhile have the same access pattern at all time) would overlook a large portion of relevant connections between domains, and thus could only detect a small number of malicious domains. Ideally, we would like domain associations to be both highly accurate and with high coverage. However, in practice it becomes quite challenging to do so because, depending on how they are collected, the scope of information in DNS data could be very limited.

In this paper, we investigate the design of effective domain association schemes based on {\em active} DNS data to enhance inference-based malicious domain detection. Active DNS data is collected by periodically querying a large pre-compiled list of domains in the Internet. One advantage of active DNS data is that it does not contain Internet activities of any real users, and could be made widely available with no privacy concerns. However, compared to other types of DNS data (e.g., passive DNS data or logs of DNS servers), the information provided by active DNS data is quite limited: only the (possibly incomplete) mapping between domains and IPs at different periods of time is available. This imposes significant challenges to reveal meaningful connections between domains. 

Belief propagation (BP), as a popular inference algorithm, has been successfully applied to analyze system and network logs and infer malicious entities. It is compelling to apply BP directly to detect malicious domains by treating the active DNS data as a bipartite graph, where on one side are domains and the other are IPs, and an edge represents the resolution of a domain to an IP. This approach implicitly builds associations between domains if they are resolved to the same IP (a ``co-IP'' relationship). Intuitively, it seems that domains hosted at the same IPs tend to be related and could be used for maliciousness inferences. However, as pointed out by Khalil et al.~\cite{GuiltyByAssociation_Khalil2016}, this is a very weak association as it fails to consider the many complicated ways that domains are deployed in the Internet. In particular, public web hosting services could cause unrelated domains to be hosted at the same pool of IPs. Malicious domain inference based on co-IP relationships thus would result in a very poor detection accuracy. 

To overcome the inherent weakness of simple ``co-IP''  associations, Khalil et al. proposes to establish explicit association between two domains if they share common IPs that do not all come from the same Autonomous Systems (AS). The number of ASs of these common IPs further indicates the strength of the association. The intuition is reasonable: even with web hosting services, it is unlikely that two unrelated domains happen to be hosted at the same set of IPs belonging to different ASs during a short period of time (say a week). The more the number of different ASs among the shared IPs is, the more likely their deployments are coordinated, and thus the more likely these domains are related. 

However, the above associations, though most of the time accurate, tend to be overly restrictive, which would result in poor coverage of malicious domains. Specifically, it largely focuses on domains that frequently change hosting environments among different ASs. Related domains that do not change hosting IPs and those that change hosting IPs within the same AS would be omitted. In fact, we notice that the majority of domains with shared IPs in active DNS data belong to this uncovered category (e.g., 82\% of domain pairs with common IPs in the data of the week of Feb. 4-10 2017 belong to this category). Meanwhile, the accuracy of associations could still be affected by web hosting services whose hosting IPs span over multiple ASs (e.g., both AS16509 and AS14618 belong to Amazon). Associations could still be incorrectly established between unrelated domains if they happen to use the same web hosting service. 

Our goal in this research is to define a new set of richer and stronger associations that both expand the coverage of domains as well as improve the detection accuracy. To expand the coverage of domains, we first explore the possibility of differentiating between dedicated hosting environments and the public hosting ones. Our intuition is that domains that share dedicated IPs are likely owned by the same entity, and hence are related, irrespective of the number of ASs. For this purpose, we identify a set of features of IPs in active DNS data, and build a classifier to distinguish these two types of different IPs. We then propose a new association scheme based on the classification of IPs. Specifically, two domains are associated if they either (i) share at least one dedicated IP, or (ii) share more than one public IP from different ASs. Our experiments show that, used within an existing path-based inference algorithm~\cite{GuiltyByAssociation_Khalil2016}, the new associations significantly expand the coverage of related domains. More importantly, the proposed new association proves to be also stronger and more meaningful as evidenced by the improvement in detection accuracy.

The path-based inference algorithm is specifically designed for DNS data analysis. It is effective but computationally expensive. To further demonstrate the strength of our domain association scheme as well as improving inference efficiency, we further investigate the effectiveness of combining our association scheme with the generic belief propagation algorithm. Through comprehensive experiments, we show that this approach significantly improves the efficiency and scalability of malicious domain inference with only minor negative impacts on detection accuracy, which suggests that such a combination would offer a good tradeoff for malicious domain detection in practice.

The rest of the paper is organized as follows. In Section~\ref{background}, we provide background material about DNS data, and the state of the art inference algorithms, namely, Belief Propagation and Path-based algorithm. Section~\ref{construction} defines and constructs domain associations. Specifically, in this section, we first provide details about our novel IP classification scheme and then use it to construct domain graphs. Section~\ref{experiment} presents the experiments to show the domain coverage and detection accuracy. We discuss detection accuracy and scalability in Section~\ref{perf}. Section~\ref{related} critically analyzes the work closely related to ours and justifies why our approach is superior. Finally, we conclude the paper in Section~\ref{conclusion}.

\section{Background}\label{background}
This section provides an overview of Domain Name System (DNS) data and the two state-of-the-art malicious domain inference algorithms, namely belief propagation and path-based.

\subsection{DNS Data}\label{dnsdata}

DNS is a hierarchical naming system that helps locate any resource connected to the Internet. In fact, it is one of the core protocol suites of the Internet. It provides a distributed database that maps domain name to record sets, such as IP addresses. DNS consists of recursive resolvers, authoritative name servers, and root name servers. Every DNS query starts from a recursive resolver, which usually resides within the local network where the query is issued. The recursive resolver  returns the IP address of the queried domain from its cache, if present, if not it queries a root server. The root server responds with a referral to a Top Level Domain (TLD) server such as \textit{com} or \textit{org} DNS server. Subsequently, through the TLD server, the recursive resolver is provided with a referral to an authoritative server, which stores the domain name to IP mapping. Finally, the DNS query is completed with a resolved IP address for the queried domain name.

DNS data can be obtained by deploying sensors in the DNS query process. Based on the location and the method of collection,  the DNS data 
can be grouped into three main categories: passive DNS data, active DNS data, and logs of DNS servers. 
The logs of DNS servers provide the richest source of information as they directly capture activities of each individual user. Passive DNS~\cite{PassiveDnsReplication_Weimer2005} captures traffic by cooperative deployment of sensors in various locations of the DNS hierarchy. For example, Farsight passive DNS data~\cite{DNSDB} utilizes sensors deployed behind DNS resolvers and provides aggregate information about domain resolutions. Active DNS data is collected by periodically querying a large pre-compiled list of domains in the Internet (e.g.,~\cite{Thales_Kountouras2016}). It is true that passive DNS data has been an invaluable source of information for detecting and mitigating malicious activities in the Internet~\cite{FromThrowAwayTrafficToBots_Antonakakis2012,GMAD_Lee2014,GuiltyByAssociation_Khalil2016}. However, in this paper, we focus on active DNS data due to the difficulty of obtaining other types of DNS data including passive DNS data and logs of DNS servers because of sensitivity of information or financial costs~\cite{DNSDB,Thales_Kountouras2016}. 

One advantage of active DNS data is that it does not contain Internet activities of any real users, and could be made widely available with no privacy concerns. However, compared to other types of DNS data, 
active DNS data offer limited information: only the (possibly incomplete) mapping between domains and IPs at different periods of time. Further, if DNS queries are issued from a limited set of hosts, the collected DNS data could be localized to the locations of those hosts and may fail to identify all IPs associated with a given domain (e.g., websites on content distribution networks). This imposes significant challenges to identify meaningful connections between domains. Another limitation is that the techniques that rely on user level features of DNS data, such as user query patterns, to detect malicious domains fail to work with active DNS data as DNS queries are not made by individual users but rather by the data collector. Despite these limitations, we show in this paper that active DNS data can be utilized to construct strong associations among domains and subsequently detect malicious domains with high accuracy.

\subsection{Belief Propagation (BP)}

BP~\cite{BP:Yedidia:2001} is an efficient approximation algorithm for solving inference problems on graphical models such as Bayesian networks~\cite{Friedman:1997:BNC} and Markov random fields~\cite{Rue:2005:GMR}. 
The algorithm was first proposed by Judea Pearl~\cite{Judea:1982:BP} in 1982. BP was initially formulated on trees, but since then it has shown to work on poly-trees and then subsequently on general graphs~\cite{Yedidia:2003:BP}.

First, we provide the inituition and high-details of BP. In BP, each node infers a final belief distribution by listening to its neighbors.
BP is an iterative message passing process. In each iteration, each node updates and passes messages to its neighbors based on the messages it received from other neighbours in the previous iteration. This process continues until all the messages converge. The final belief of each node is calculated based on the final messages. The high-level idea is that given a small set of labeled nodes, BP infers the labels of the rest of the nodes in the graph. The propagation of belief also depends on either the {\em homophily} ("birds of a feather flock together") or the {\em heterophily} ("opposites attract") property of a given network. The network graphs we are dealing with in this work are known to demonstrate certain homophily properties. That is, if a node is surrounded by many malicious nodes than benign nodes, that node is likely to be malicious as well.

BP has already been applied in the field of malicious domain detection~\cite{Jiang:2010:ISA,DetectingMaliciousDomainsViaGraphInference_Manadhata2014}. For example, Manadhata et. al~\cite{DetectingMaliciousDomainsViaGraphInference_Manadhata2014} apply BP over host-domain bipartite graphs, which represent hosts querying domains, in order to discover new malicious domains based on known malicious and benign domains as ground truth. They collected passive DNS data from HTTP proxies deployed in a global enterprise. The intuition behind their associations is that if a host access a malicious domain, it is likely to be infected or compromised. Further, the domains queried by this infected host are likely to be malicious as well. In fact, Manadhata et. al empirically show that the belief propagation algorithm achieves high detection rates of malicious domains with low false positive rates. In this work, we adopt the BP implementation in~\cite{DetectingMaliciousDomainsViaGraphInference_Manadhata2014}.

Now we are going to take a closer look at the BP algorithm for the benefit of the readers who are not familiar with BP. Specifically in BP, each node tells each neighboring node what its belief on each possible states (i.e. the marginal probability of being in each state) based on its initial belief of itself and the belief of the rest of its neighboring nodes. For example, if most of the neighbors of a node are malicious, that node informs other neighbours that it believe that they are more malicious than benign. BP iteratively propagates the beliefs of a small set of nodes with known beliefs throughout the graph until convergence.

Formally, given an undirected graph $G = (V, E)$, where $V$ and $E$ are the set of nodes and edges, respectively. Each node $v_i\in V$ is modeled as a discrete random variable $X_i$, where $i\in \{1,n\}$ and $n=|V|$. Each random variable can be in one of the finite states in $S$, that is $s_j\in S$, where $j\in\{1,m\}$ with $m$ finite states. For example, in our problem setting, each node represents a domain or an IP and each discrete random variable can be in either malicious or benign state. This graphical model defines a joint probability distribution $P(X_1, X_2,\dots, X_n)$ over the nodes in $G$. By "belief", we simply mean the marginal probability being in each possible state. The exact inference algorithm computes the marginal probability distribution $P(X_i)$ of each discrete random variable $X_i$ given a set of observed/labeled nodes, ground truth in our case. As shown in Formula~\ref{marginal}, the marginal probability of a given node, say $X_j$, can be computed by taking the sum over all possible states of all the other nodes in the graph. The complexity of this equation is exponential in the number of nodes. As a scalable solution, we adopt the BP approximation provided in ~\cite{DetectingMaliciousDomainsViaGraphInference_Manadhata2014}, which computes approximate marginal probabilities with quadratic complexity in the number of nodes, that is $O(V^2)$ in the worst case.

\begin{equation}\label{marginal}
p(X_j) = \sum_{X_i, i \in\{1,n\}\setminus j} p(X_1, X_2, X_3, \dots, X_n)
\end{equation}

As mentioned earlier, the labeled nodes represent the ground truth. They simply provide us the prior probability of being in each possible state. Usually, $\phi_i(s_j)$ is used to denote the {\em prior probability} of node $i$ being in state $s_j$. BP algorithm infers the beliefs, $b_i(s_j)$ for each node $i$ by computing the marginal probability for each possible state $s_j$, $j\in\{1,m\}$. The statistical properties between any two connected nodes $i$ and $j$ is captured by the {\em edge potential} function denoted by $\psi_{ij}(s_i, s_j)$, where nodes $i$ and $j$ are in states $s_i$ and $s_j$ respectively. As in our application, if each node can be in one of the two states, benign or malicious, $\psi_{ij}(s_i, s_j)$ can be represented by a $2\times 2$ matrix, where each represents a possible state of the two edge nodes.

In each iteration, each node $i$ passes a message vector, $m_{ij}$, of size $|S|$ to each neighbor $j$ and $m_{ij}(s_r)$ denotes the message from $i$ to $j$ for state $s_r$. It is computed as shown in Formula~\ref{sumproduct}.

\begin{equation}\label{sumproduct}
m^{t}_{ij}(s_r)  = \sum_{s_p \in S} \left(\phi_{i}(s_p)  \psi_{ij}(s_p,s_r) \prod_{k \in \mathcal{N}_i/j} m^{t-1}_{ki} (s_p)\right)
\end{equation}

where $t$ is the current iteration count, $m^t$ is the message in iteration $t$, $m^{t-1}$ is the message in the previous iteration, and $\mathcal{N}$ is the set of neighboring nodes of $i$. The message $m^t_{ij}(s_r)$ captures node $i$'s belief of node $j$ being in state $s_r$ based on the information node $i$ received from its neighbors during the previous iteration.

The algorithm stops when the difference between messages from two consecutive round for all edges is below a predefined threshold value. BP is proved to be always converging for trees, but it may or may not converge for graphs with loops. Having said that, in practice BP has been shown to produce reasonably accurate results for even graphs with loops. Usually, BP algorithm is run until one of the following conditions are met: (1) BP converges to the specified threshold value or (2) A predetermined number of message iterations are completed.

Once BP stops, Formula~\ref{belief} shows how the final belief is computed for each node.

\begin{equation}\label{belief}
b_i(s_r)  = C\phi_{i}(s_r)\prod_{k \in \mathcal{N}_i} m_{ki} (s_r)
\end{equation}

where $C$ is normalizing constant such that $\sum_{s_p\in S} b_i(s_p) = 1$.

\subsection{Path based Algorithm}
Khalil et al.~\cite{GuiltyByAssociation_Khalil2016} designed a path-based malicious domain detection algorithm over a domain graph, which is built from a domain resolution graph (which domains are being hosted at which IPs). A domain graph comprises nodes representing domains and weighted edges that represent strengths of direct associations between domains. The algorithm further computes indirect associations between pairs of domains. Specifically, the strength of a path between two nodes $u$ and $v$ is defined as the multiplication of the weights of the edges on the path. The association between $u$ and $v$ is defined as the weight of the strongest path between them. Given a set of known malicious domains, the maliciousness of a domain is then inferred as an exponentially decaying weighted sum of its associations with all known malicious ones. 

More specifically, let $S$ be the set of known malicious domains, also called seed. Given a domain $u$, denote $M(u)$ as the list $(assoc(s_1, u), \dots, assoc(s_n, u))$, where $s_i \in S$ and $assoc(s_i, u) >= assoc(s_{i+1}, u)$, for $i = 1, \dots, n-1$. In other words, $M(u)$ is a sorted list of the associations of $u$ to each of those in the seed. The malicious score of $u$ given $S$ is then defined as:

\begin{equation}\label{association}
\begin{split}
mal(u,S) & =  assoc(s_1, u) + \\
         & \Big(1-assoc(s_1,u)\Big)\sum_{i=2,\dots, n}\frac{1}{2^{i-1}}assoc(s_i,u))
\end{split}
\end{equation}

Intuitively, the largest association between $u$ and a known malicious domain contributes the most to the maliciousness of $u$. This is further enhanced with its association with other domains in the seed in an exponential decay manner. This design is to capture two intuitions of malicious domain inferences. First, a strong association with even a single known malicious domain would be convincing evidence of a potential malicious domain. Second, weak association with multiple known malicious domains cannot be easily accumulated to form strong evidence of a domain's maliciousness, because weak associations may happen in many legitimate network management scenarios.

In this paper we explore using both BP and the path-based algorithm for malicious domain inferences based on our association scheme and show how our proposed domain association scheme could help improve detection accuracy and domain coverage.

\section{Defining and Constructing Domain Associations}\label{construction}

In this section, we first provide the details of IP classification as a prerequisite for defining associations. Then, we define our associations and the domain graphs generated from them. Finally, we conduct experiments to evaluate the detection accuracy and the domain coverage of our associations.

\subsection{IP Classification}\label{IP_Classification}
As mentioned in Section~\ref{introduction}, we classify the IPs in domain-IP graphs into {\em public} and {\em dedicated} as a basis for defining a new set of stronger and more meaningful domain associations. We define public domain-hosting IPs, public IPs for short, as those IPs that are publicly available for domain hosting and consequently, host domains from multiple unrelated entities (e.g. Amazon Public IPs~\cite{AWSPublicIP}, Ephemeral external IPs in Google Cloud Platform~\cite{GooglePublicIP}). In contrast, dedicated domain-hosting IPs, dedicated IPs for short, are defined as those IPs that are exclusively used to host domains of the same entity for a certain time interval (e.g. Amazon Elastic IPs~\cite{AWSElasticIP}, IPs of most universities and government institutions). 

Naturally, the first option to identify public IPs is to check in publicly available APIs from hosting providers~\cite{GooglePublicIPAPI} or static cloud IP lists compiled by hosting providers~\cite{AWSPublicIPList} and third parties~\cite{AzurePublicIPList}. However, such APIs may have rate limits of checking or such static lists may often compiled manually and cannot keep up with domain hosting dynamics. Additionally, such tools usually have limited coverage of IPs. Most importantly, they may not capture the notions of public and dedicated IPs as we define in this study. For example, Which-Cloud~\cite{WhichCloud} not only have limited coverage of cloud hosting IPs, but also do not differentiate between public and dedicated IPs. More specifically, Which-Cloud classifies Amazon Elastic IPs~\cite{AWSElasticIP} as public. However, these IPs are usually allocated to one entity for a certain time interval, and hence, are considered dedicated according to our definition. For example, consider the IP "35.160.237.41" from the domain graph of the week of Feb. 4-10 2017. Which-Cloud classifies this IP as public since it belongs to Amazon AWS IP pool. However, we classify this IP as dedicated, even though it belongs to a public hosting organization. By consulting DNS databases as well as VirusTotal~\cite{VirusTotal}, we verify that this IP is used to host five domains during our study period: spoonerrisk.com, spoonerinc.net, spoonermai.com, medadmin.com and thomasproductiontrainingmanuals.com, all of which belong to the same entity, Spooner Inc. and its subsidiaries. 

In the following, we walk through the details of our solution to classify IPs into dedicated and public according to our definition of the two classes.

\textbf{Attribute Selection.} The accuracy of classification depends on the selection of the right attributes of DNS resolved IPs. After studying many IP attributes, we selected 7 attributes that collectively differentiate dedicated IPs from public ones. As shown in Table~\ref{tab:attributes}, we categorize them into two groups: (i) domain based attributes and (ii) IP block based attributes. The three attributes in the first group are: the number of fully qualified domain names (FQDNs) (e.g., www.foo.example.com is a FQDN), the number of second level domains (2LDs) (e.g., example.com), and the number of third level domains (3LDs) (e.g., foo.example.com), which an IP hosts during a certain time period (e.g., one week). Attributes 4-6 (Table~\ref{tab:attributes}) of the second group are similar to attributes 1-3, but computed over all the IPs, in the selected data set, which belong to the same /24-subnet of the IP under consideration. Attribute 7 is the number of IPs in the /24-subnet of the IP under consideration, that has been used to host domains in the selected data set.   


\begin{table}[!t]
\renewcommand{\arraystretch}{1.7}
\newcounter{magicrownumbers}
\newcommand\rownumber{\stepcounter{magicrownumbers}\arabic{magicrownumbers}}
\caption{Selected IP Attributes for IP Classification}
\label{tab:attributes}
\centering
\begin{tabular}{|c|c|c|l|}
\hline
Attributes Set & \# & \multicolumn{1}{|c|}{Attributes Name for IP} \\
\hline
\multirow{3}{*}{Domain based attributes} & \rownumber & \multicolumn{1}{|l|}{\# of FQDNs} \\
\cline{2-3}
& \rownumber & \multicolumn{1}{|l|}{\# of second level domains} \\
\cline{2-3}
& \rownumber & \multicolumn{1}{|l|}{\# of third level domains} \\
\hline
\multirow{4}{*}{IP block based attributes} & \rownumber & \multicolumn{1}{|l|}{\# of FQDNs in its /24 IP block} \\
\cline{2-3}
& \rownumber & \multicolumn{1}{|l|}{\# of second level domains in its /24 IP block} \\
\cline{2-3}
& \rownumber & \multicolumn{1}{|l|}{\# of third level domains in its /24 IP block} \\
\cline{2-3}
& \rownumber & \multicolumn{1}{|l|}{\# of IPs in its /24 IP block} \\
\hline
\hline
\end{tabular}
\end{table}

Domain based attributes capture statistics about the domains that each IP is associated with. 
Intuitively, a public IP has many second level domains, whereas a dedicated IP has less. Web-hosting services and cloud providers host many domains under the same IP block. That is, such public hosting providers usually allocate complete blocks or large portion of the same IP block for public hosting, which results in a large number of domain resolutions within such blocks. On the other hand, dedicated hosting usually uses a fewer number of domains within their IP blocks. The attributes of the second group help to correctly classify dedicated IPs that happen to host a relatively large number of domains (e.g., a big company and its subsidiaries), which otherwise may be misclassified based on the first group of features.  

\textbf{Ground Truth Collection.} To train any classifier, one needs to first compile a ground truth of labeled data. We use the following process to collect a ground truth consisting of both dedicated IPs and public IPs. We manually verify these IPs, by checking whether the domains hosted at each of these IPs belong to the same entity using WHOIS database~\cite{Whois}.

\begin{itemize}
\item 

     \textbf{Dedicated IP ground truth collection}: Intuitively, it is less likely for a university or a government website to share IPs with domains from other different organizations. We compiled a set of organizations that are likely to use dedicated IP hosting for their domains including known universities from different countries (e.g., MIT, Purdue, Tsinghua University, Oxford university), governmental organizations (e.g., U.S. Dept. of Health and Human Services, Sandia National Laboratories), military related organizations (e.g., US Coast Guard Home, U.S. Air Force, United States Army), and private organizations (e.g., Google, Oracle, AliBaba, NJEDge.Net, Inc., Merit Network Inc.). Such communities of organizations usually use certain top level domains such as, .gov, .edu, .ml, and .com. We then compile the set of IPs that are used to host domains from these communities. We check all the ASs of the IPs in a one-week active DNS data and search for keywords such as ``University'' and ``Government'' to identify research institutes and government agencies. For other types of organizations, we simply randomly select IPs from the active DNS data. Then, we manually check if these IPs are historically only used to  host domains belonging to specific organizations. For example, for purdue.edu, the AS of its IPs is indeed ``PURDUE - Purdue University, US'', and the IP ``128.210.7.200'' has been used to host only the domain purdue.edu at least since 2014. We conduct further manual verification when the AS of an IP is not directly linked to the domain. For example, the AS of the IP of tsinghua.edu.cn is "China Education and Research Network Center", which does not seem directly related to Tsinghua University. However, further investigation shows that Chinese universities indeed get their dedicated IPs from this AS to host their domains. 

\item \textbf{Public IP ground truth collection}: Similar to the dedicated IP ground truth, we first select a set of second level domains of popular web-hosting and cloud provider domains, such as Amazon AWS, bulehost.com and ehost.com. Then, we identify FQDNs belonging to these 2LDs from our active DNS dataset and the resolving IPs are marked as candidate public IPs. However, some of the IPs in this list may be assigned for dedicated use. That is, some cloud IPs may be reserved and exclusively used by a single entity for a certain period of time (e.g. Amazon Elastic IPs~\cite{AWSElasticIP}), and hence may not qualify as public IPs as defined above. Therefore, we manually check domains hosted at each candidate public IPs and verify whether they are owned by the same entity based on their WHOIS records, the length of continuous duration of domain IP mapping (e.g., using the DNS TTL of domain-IP resolution), and possibly the contents of these websites.
%

\end{itemize}


\begin{figure}[!t]
\centering
\includegraphics[width=3.5in]{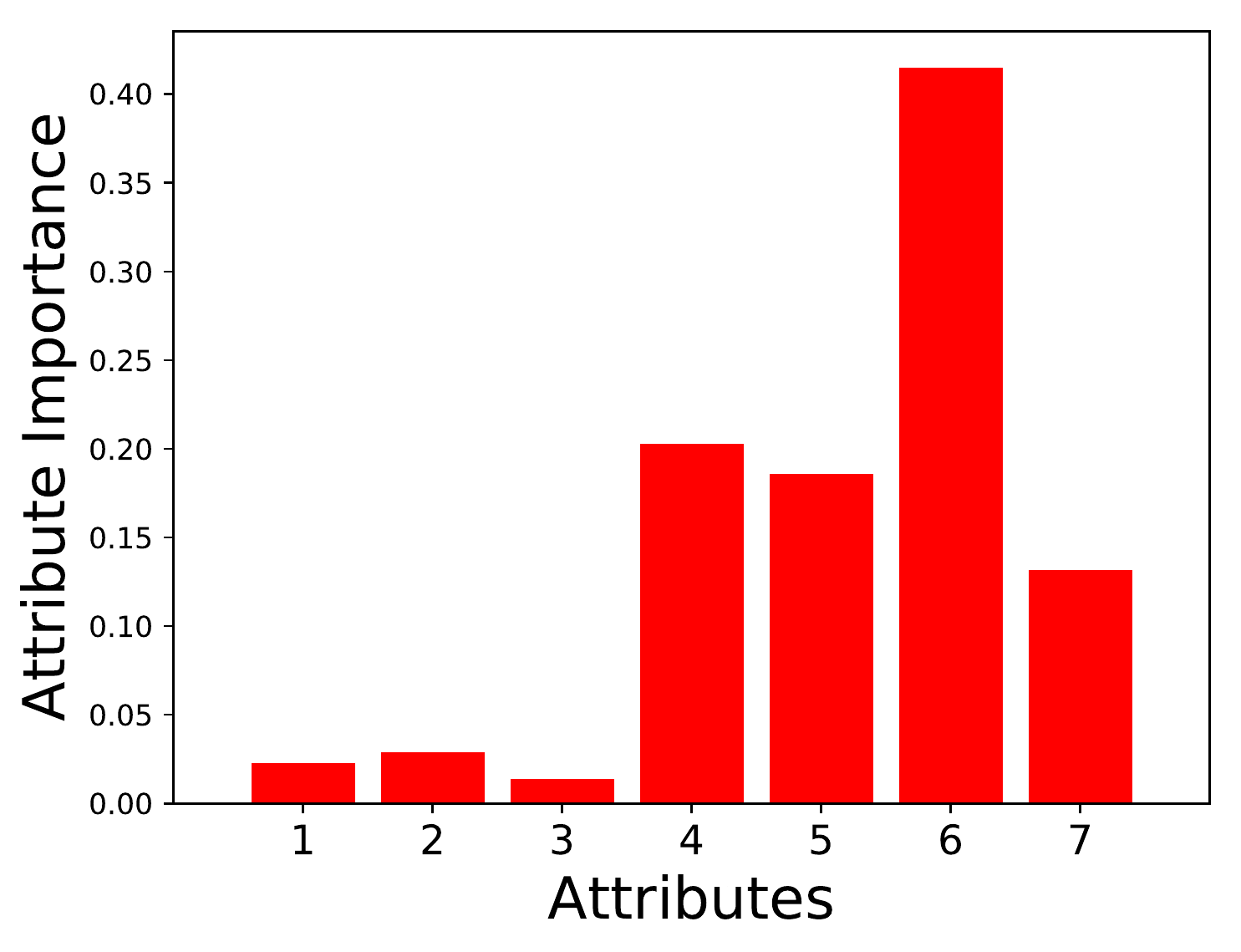}
\caption{Attribute Importance of IP classification}
\label{fig:attributeimportance}
\end{figure}

\textbf{Classification Algorithm.} We have tried different classifiers including decision trees, SVM, and random forest and found that semi-supervised random forest is the most suitable classifier for our IP classification problem \cite{Leistner:2009:SSLRandomForest}. Semi-supervised random forest is a machine learning technique where it uses labeled data (seed) as well as unlabeled data to do the classification. We design a random forest classifier based on the aforementioned 7 attributes extracted from the active DNS dataset. The classifier is implemented using the Python scikit-learn package \cite{sklearn}. Notice that random forest is an iterative algorithm. In each round, it uses the seed to classify a set of IPs from the unlabeled data set and adds it to the input seed. The new seed and the remaining unlabeled domains are then used in the second round, and so on until all the unlabeled set is labeled. The classification procedure is shown in Algorithm 1. After each classification round, we revise the seeds based on the two thresholds, $confidence\_thresh_{public}$ and $confidence\_thresh_{dedicated}$. We set the thresholds $confidence\_thresh_{public}$ and $confidence\_thresh_{dedicated}$ to $0.5$ and $0.9$ respectively in order to obtain a classification accuracy above $98\%$. The subsequent iterations uses the manually collected seed as well as the already labeled IPs as the new seed. 
The algorithm converges when the two output sets, $IP_{public}$ and $IP_{dedicated}$, remain unchanged. We observe that random forest classifier converges in $10$ to $11$ rounds for the active DNS dataset we use in this study. Figure~\ref{fig:attributeimportance} shows the importance of each attribute after convergence. Out of the $7$ attributes, IP block based attributes clearly have more significance in the classification compared to domain based attributes. While some dedicated IPs (e.g., IPs of big organizations) may exhibit similar domain based attributes to public IPs, dedicated IP blocks are highly unlikely to exhibit similar attributes to public IP blocks, which makes IP block based attributes more distinctive. In other words, it is more likely for IPs in public blocks to have consistent domain attributes compared to IPs in dedicated blocks, which results in different block based domain attributes.

The /24-subnet for the IP classification is chosen based on the accuracy of the classifier. We evaluated the accuracy of the classifier for different subnet sizes from 12 to 30. It is observed that when the subnet sizes are either large, i.e. close to 12 or small, i.e. close to 30, the accuracy of the classifier falls below 98\%. Based on the accuracy, we identified several candidate subset sizes including 18, 20, 22, 24 and 26. As explained later, we selected /24-subnet out of the above candidate subnets as it results in the lowest false positive rate when detecting malicious domains. Intuitively, large blocks are most likely shared by multiple entities, which leads to wrong associations between domains that are not owned by the same entity and hence results in high false positives. On the other hand, small blocks may split domains owned by the same entity, which may result in missing important associations among domains owned by the same entity.

\begin{algorithm}[t]
\label{alg:classification}
\SetAlgoLined
\caption{IP Classification Algorithm} 
\SetKwInOut{Input}{Input}
\SetKwInOut{Output}{Output}
\Input{$IP_{p\_labeled}$, public IPs in the seed\\
       $IP_{d\_labeled}$, dedicated IPs in the seed\\
       $IP_{unlabeled}$, DNS resolved IPs to be classified\\}
\Output{$IP_{public}$, IPs classified as public including seed \\
        $IP_{dedicated}$, IPs classified as dedicated including seed\\}

\BlankLine
\BlankLine
$IP_{public}, IP_{dedicated} = IP_{p\_labeled}, IP_{d\_labeled}$ \\
\While{True}{
    model = train\_classifier\_module($IP_{public}$, $IP_{dedicated}$) \\
    S = predict\_unknown\_ip(model, $IP_{unlabeled}$) \\
    \For{each element in S}{
        \# IP address \\
        $IP$ = $element.ip$ \;
        $score_{public}$ = $element.public\_confidence\_score$ \;
        \# dedicated confidence score \\
        $score_{dedicated} = 1 - score_{public}$ \;
		\uIf{$score_{public} > confidence\_thresh_{public}$}{
			Move $IP$ from $IP_{unlabeled}$ to $IP_{public}$\;
		}
		\uElseIf{$score_{dedicated} > confidence\_thresh_{dedicated}$}{
			Move $IP$ from $IP_{unlabeled}$ to $IP_{dedicated}$\;
        }
	}
	\# convergence \\
	\If{$|IP_{public}|$ is not changed and $|IP_{dedicated}|$ is not changed}{exit while\;}
}
\BlankLine
\For{each $IP$ in $IP_{unlabeled}$}{
	Move $IP$ to $IP_{public}$ \;
}
\Return $IP_{public}$, $IP_{dedicated}$
\vspace{2em}
\label{alg:classification}
\end{algorithm}

\subsection{Construction of Domain Graphs}\label{construction}

IP classification is motivated by the need to define a new set of domain associations that strike the right balance between domain coverage and detection accuracy. As mentioned earlier, a weak or irrelevant association would result in high false positives; meanwhile, an overly restrictive/strong association will overlook many potentially malicious domains as they do not possess such a strong association with known malicious ones. In other words, many such domains would be filtered out before an inference algorithm could be applied. Our key observation is that dedicated IPs host domains that belong to the same entity, and hence are related. On the other hand, public IPs host domains from many unrelated entities, and therefore, sharing a public IP is not a conclusive evidence of a good relationship among domains. In the latter case, we utilize the traces left by the evading behavior of malicious domains to create associations among them. More specifically, some owners of malicious domains frequently change their hosting environment to evade detection, which creates intrinsic relations among them. We use here the same heuristic that is used in~\cite{GuiltyByAssociation_Khalil2016} to capture such behavior, that is, the number of common ASs in which domains are hosted during a certain time period. 

We define two types of new domain associations based on the outcome of the IP classifier: 
\begin{itemize}

    \item The first association type utilizes the common dedicated IPs as well as the common ASs between domains. More specifically, two domains are associated if they: share either (i) at least one dedicated IP, which we call the \textit{dedicated association rule}; or (ii) more than one public IP from more than one AS, which we call the \textit{public association rule}. The dedicated association rule is motivated by the fact that sharing even a single dedicated IP is a good evidence of a relation since, by definition, dedicated IPs host domains that belong to the same entity. The public association rule is triggered by the observation that some malicious domain owners avoid using dedicated IPs as they may be easily identified and blocked. At the same time, they frequently move across different hosting environments to avoid detection and blocking of the domains themselves.\\

The strength of the first type of associations depends on the number of shared dedicated and public IPs. Specifically, given a pair of domains $d_1$ and $d_2$ that share a set $I$ of resolved IPs, let $IP_d$ denote the set of resolved dedicated IPs in $I$, $IP_u$ is the set of resolved public IPs in $I$, while $AS(IP_u)$ denote the set of ASs that the resolved public IPs in $I$ belong to. We define the association weight between two domains $d_1$ and $d_2$, $w(d_1,d_2)$, as

\begin{equation}\label{new_weight}
\begin{aligned}
w(d_1, d_2) = \begin{cases}
                    1 - \frac{1}{n + 1}; & \text{if $d_1 \neq d_2$}\\
                    1 \qquad \quad; & \text{if $d_1 = d_2$}\\
              \end{cases}\\\\
where \quad n = 2|IP_d| + |AS(IP_u)| - 1
\end{aligned}
\end{equation}

Note that the weight is in $[0,1]$ and the weight of the association between the domain and itself is 1. Formula~\ref{new_weight} captures four intuitions: (1) A minimum of one dedicated IP or two common ASs are required to establish an association, (2) sharing a dedicated IP is stronger than sharing AS, (3) the more the number of ASs and dedicated IPs are, the stronger the association, and (4) the size of ASs and dedicated IPs set has a diminishing return of strength.

    \item The second type of associations is similar to the first type with only one difference. It relaxes the dedicated association rule by using the shared /24-subnets instead of the shared dedicated IPs. More specifically, we replace the dedicated association rule with a new rule, dubbed \textit{relaxed association rule}. In the relaxed association rule, two domains are associated if they are resolved to dedicated IPs that belong to the same /24-subnet. The strength of this type of associations follows the same intuition as that in the first type of associations with some changes in Formula~\ref{new_weight}. More specifically, we replace $2|IP_d|$ in Formula~\ref{new_weight} with $|IP_{d1}| + |IP_{d2}|$, where $IP_{d1}$ and $IP_{d2}$ are the set of dedicated IPs in the shared /24-subnets to which domains $d_1$ and $d_2$ are resolved, respectively. For example, assume that $d_1$ is resolved to $ip_1$ and $ip_2$ in $subnet_1$, $ip_3$ in $subnet_2$, and $ip_4$ in $subnet_3$; while $d_2$ is resolved into $ip_2$ in $subnet_1$, $ip_5$ in $subnet_2$. The number of shared subnets between $d_1$ and $d_2$ is $2$ ($subnet_1$ and $subnet_2$), $|IP_{d1}| = 2+1 = 3$, and $|IP_{d2}| = 1+1 = 2$. 

\end{itemize}

\begin{figure*}
  \centering 
  \includegraphics[width=0.9\textwidth]{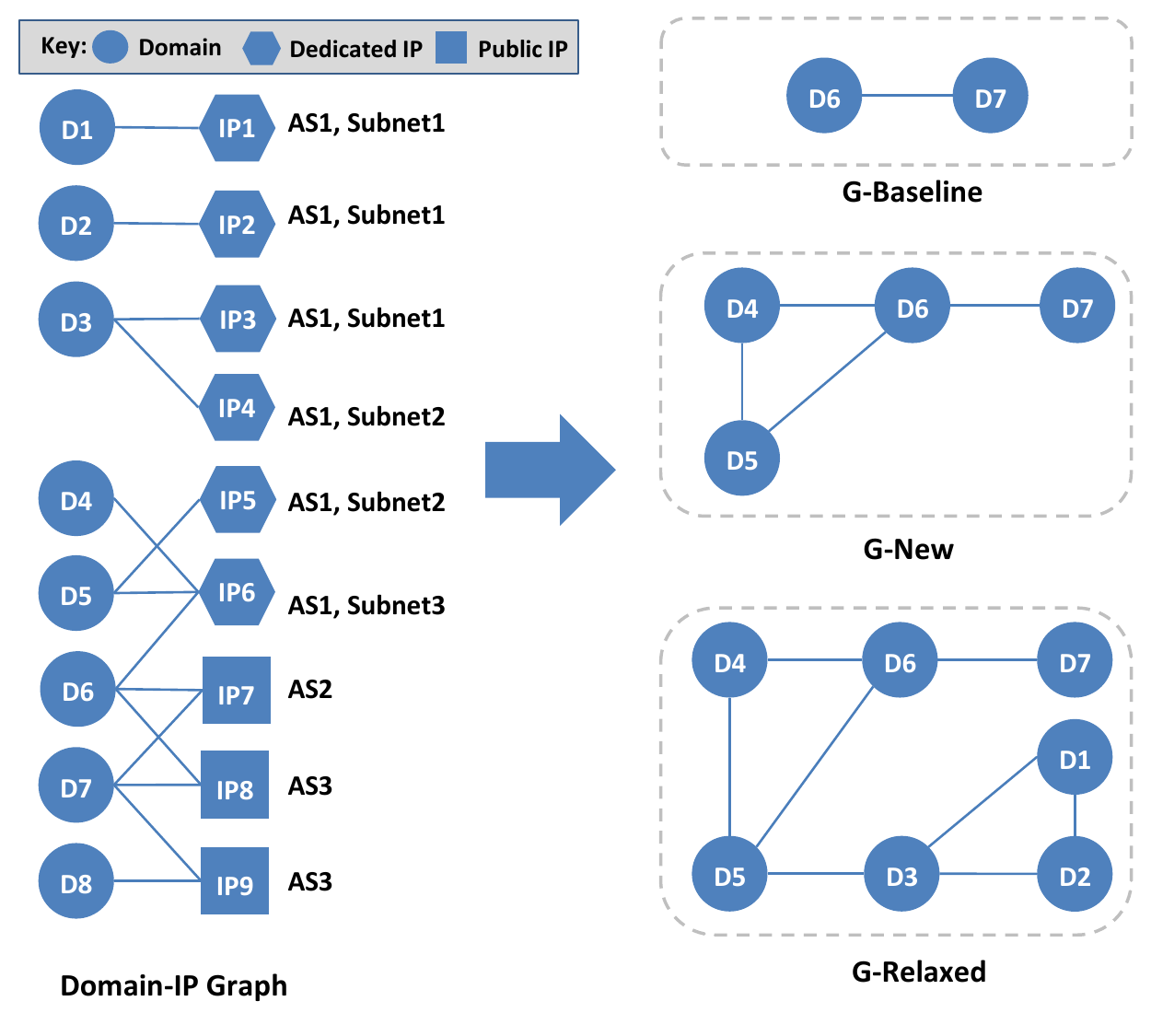}
  \caption{An example domain-IP graph and its corresponding domain graphs}
  \label{graph_construction}
\end{figure*}

We build two domain graphs based on the aforementioned two types of associations. The first graph captures the first type of associations and is dubbed, \textit{G-New} to differentiate it from the domain graph in~\cite{GuiltyByAssociation_Khalil2016}, which we call \textit{G-Baseline}. The second graph captures the second type of association and is dubbed, \textit{G-Relaxed}. Figure~\ref{graph_construction} shows an example of an original domain-IP graph and the three constructed domain graphs, \textit{G-Baseline}, \textit{G-New}, and \textit{G-Relaxed}.

\section{Domain Coverage Versus Detection Accuracy}\label{experiment}

In this section we study the impact of the different types of associations on domain coverage and detection accuracy. We first explain the dataset and the ground truth used, then we present the results.


\indent{\bf Active DNS Dataset.}
Our experiments are performed on the active DNS datasets made available by the system called Thales by the Georgia Institute of Technology's Active DNS project~\cite{Thales_Kountouras2016,ActiveDNSProjectSite}. 
Thales scans DNS using a set of seed domain list compiled from multiple sources, including public blacklists (e.g.~\cite{BlackholeDNS,ZeusDomainsBlocklist,MalwareDomains}), 
the Alexa list~\cite{Alexa}, the Common Crawl dataset~\cite{CommonCrawl}, the domain feed from an undisclosed security vendor, and the zone files for the TLDs consisting of com, net, biz and org. We observe that 
Thales provides a complete coverage of domains in its seed list roughly in $3$ days.
Cyber criminals in general make extensive use of short lived random disposable domains to carry out malicious activities. Domains with long term malicious activities are likely to be identified and blocked. In general, long-lived domains are likely to be benign. Even though sometimes such benign domains may get compromised, their administrators eventually clean and regain control of them. Therefore, we focus our analysis on new domain-IP mappings that are first observed in a certain time period.
Similar to ~\cite{GuiltyByAssociation_Khalil2016}, we set the study period to one week, which is long enough for Thales to crawl all the domains in its seed, and meanwhile not too long, thus minimizing long-lived domains included.
We conduct experiments over two time periods: (i) Feb. 4-10, 2017 (week-1), and (ii) Feb. 20-26, 2017 (week-2). Each dataset comprises a list of $<domain,IP>$ tuples of domains and the hosting IPs. Each dataset is represented by a bipartite graph with domains on one side and IPs on the other. An edge is created for each $<domain,IP>$ tuple in the dataset. The bipartite graph is dubbed domain resolution graph.

\begin{figure}[!t]
\centering
\includegraphics[width=3.5in]{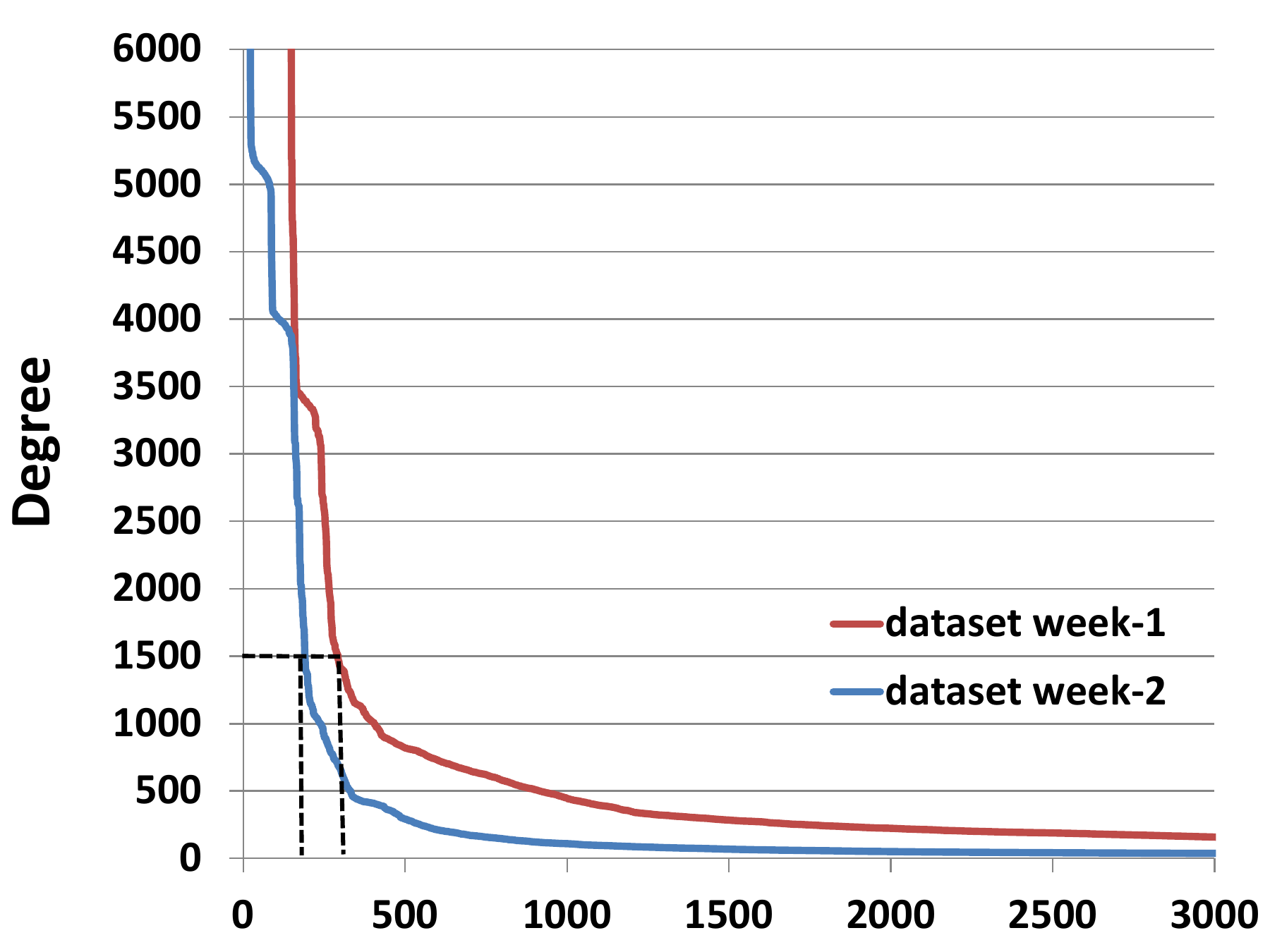}
\caption{Degree distribution of IP nodes in domain resolution graphs. Only 3000 IPs with the highest degrees are shown in the figure}
\label{fig:ip-distribution}
\end{figure}

Intuitively, web-hosting services, cloud providers and content delivery network (CDN) may host many unrelated domains under one or several IP addresses. For example, two domains hosted by the same IP in Aamzon Web Service (AWS) (or CloudFlare, Akamai) could belong to different owners. One domain being malicious does not imply that the other one is likely to be malicious. An efficient heuristic approach to fix this problem is to exclude the "popular" IPs, which host more than $t$ domains in a certain period, from the domain resolved data~\cite{GuiltyByAssociation_Khalil2016}. Figure~\ref{fig:ip-distribution} shows the degree distribution of IPs in the original domain resolution graphs of the two dataset (week-1 and week-2). The $x-axis$ is the accumulation of the numbers of IPs sorted based on their degrees, while the $y-axis$ axis shows the corresponding degrees. For example, the point (300, 1500) in Figure~\ref{fig:ip-distribution} means that there are 300 IPs each hosts 1500 domains. The IP degree distribution shows us that a small group of IPs host most of the domains. Empirically, we set $t$ to be 1500, where only 292 and 190 IPs respectively are excluded from the domain resolution graphs of the two datasets. It is a negligible percentage (0.05\% and 0.09\% respectively) of the total IPs in the original datasets.

\begin{figure}[!t]
\centering
\includegraphics[width=3.5in]{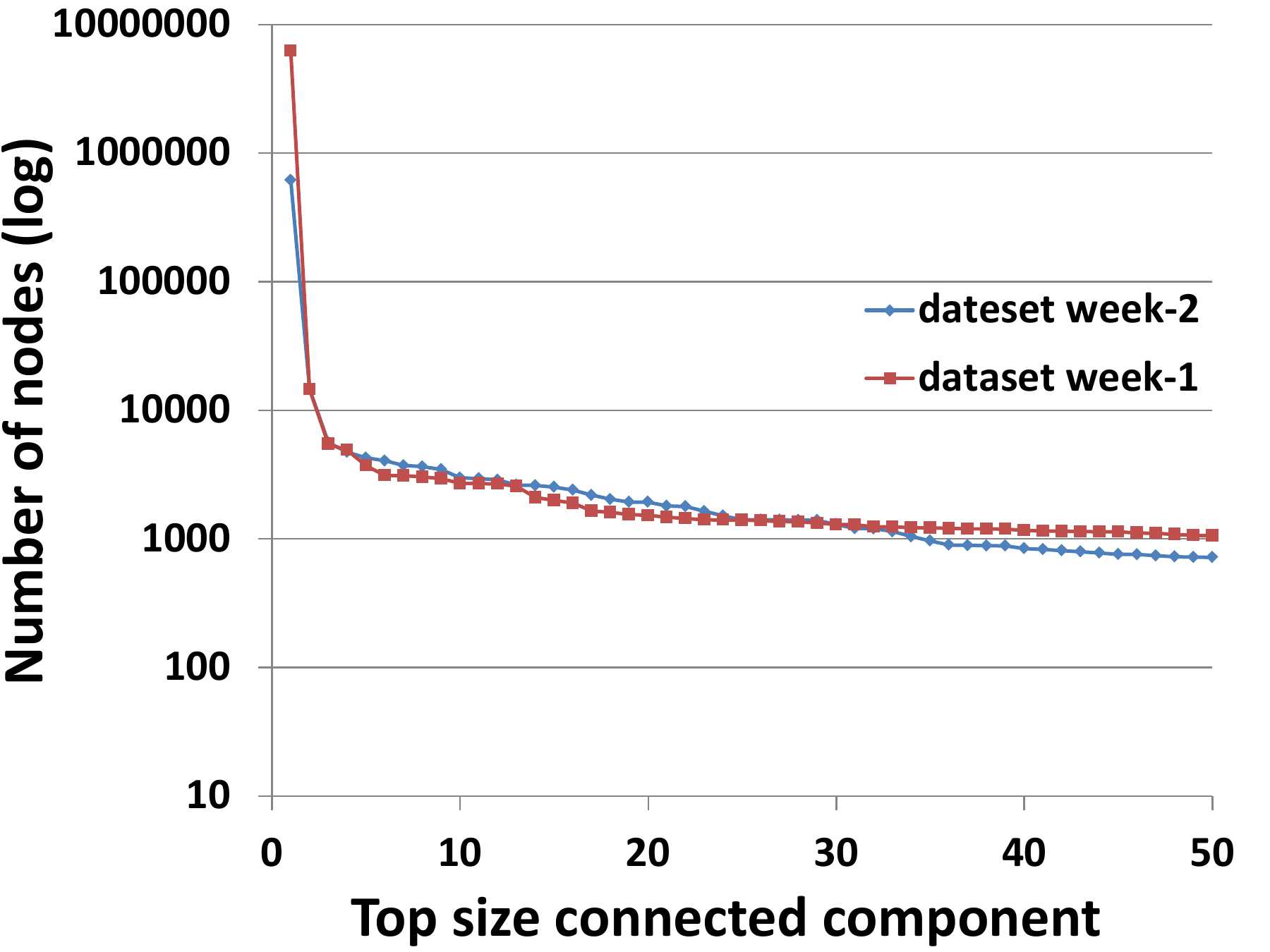}
\caption{Distribution of connected component sizes in domain resolution graphs for the two datasets. Only 50 top size connected components are shown in this figure}
\label{fig:cc-distribution}
\end{figure}

Another important property of our datasets is the number and size of connected components in the domain resolution graph. The size of connected components in the domain resolution graphs has a significant effect on the malicious score computation time, especially for message passing approaches, such as BP algorithm. Figure~\ref{fig:cc-distribution} provides the distribution of the number of nodes (domains and IPs) of the top connected components in the domain resolution graphs of the two datasets. It roughly follows a power law distribution based on the logarithmic scale. An efficient heuristic to reduce the computation time of both domain graph generation as well as inference algorithm execution is to operate on each connected component separately instead of the whole graph. 

\indent{\bf Ground Truth Collection.}
Our malicious ground truth is collected through McAfee SiteAdvisor~\cite{McafeeSiteAdvisor}. It labels each domain into one of the four categories: \textit{safe}, \textit{caution}, \textit{warning} and \textit{unknown}. The label ``safe'' indicates that a domain is benign, whereas ``caution'' is used to show that a domain may bring a minor risk. The label ``unknown'' means that SiteAdvisor does not have sufficient information to categorize a domain. Finally, the label ``warning'' indicates that a domain has a major risk of being malicious. In our experiments, we use ``warning'' domains as our malicious ground truth. 
For benign ground truth, we follow previous research in this area~\cite{DetectingMaliciousDomainsViaGraphInference_Manadhata2014,GuiltyByAssociation_Khalil2016}, and use Alexa top 1 million domains~\cite{Alexa}, in addition to the ``safe" domains from SiteAdvisor.

\indent{\bf Empirical Evaluaton.} We conduct experiments to study the impact of different types of associations on domain coverage and detection accuracy. We use the association defined in~\cite{GuiltyByAssociation_Khalil2016} as a baseline. Table~\ref{tab:graph_size} lists the sizes of the domain graphs generated from our two types of associations, as well as the baseline graph for the two active DNS datasets (week-1 and week-2). The table clearly shows that both of our proposed associations significantly expand the domain coverage. For example, \textit{G-New} is almost 10 times the size of \textit{G-Baseline} in the active DNS data of the week of Feb. 4-10, 2017. This is mainly because the associations defined in~\cite{GuiltyByAssociation_Khalil2016} are overly restrictive. Specifically, associations are established only between domains that frequently change hosting environments among different ASs. Domains that do not change hosting IPs and those that change hosting IPs within the same AS are not considered. For example, in week-1 (Table~\ref{tab:graph_size}), more than 82\% of the domains that share IPs satisfy only the dedicated association rule, and hence are completely ignored in \textit{G-Baseline}. The table also shows that our second type of associations provides even larger domain coverage compared to the first type. For example, \textit{G-Relaxed} is about 2.7 times the size of \textit{G-New} in week-1. This is due to the relaxed association rule which generates a super-set of the associations generated by the corresponding dedicated association rule. Specifically, two domains that resolve to different dedicated IPs in the same /24-subnet are associated in \textit{G-Relaxed} but are not associated in \textit{G-New}.

\begin{table}[htbp]
\renewcommand{\arraystretch}{1.7}
\caption{Sizes of Domain Graphs}
\label{tab:graph_size}
\centering
\begin{tabular}{|c|c|c|c|}
\hline
\multirow{2}{*}{Dataset} & \multicolumn{3}{|c|}{\# of Domains in domain graph} \\
\cline{2-4}
& G-Baseline & G-New & G-Relaxed \\
\hline
week-1 (Feb. 4-10, 2017) & 3,980 & 39,604 & 106,222 \\
\hline
week-2 (Feb. 20-26, 2017) & 2,449 & 18,779 & 43,661 \\
\hline
\end{tabular}
\end{table}


We conduct extensive set of experiments to infer malicious domains by applying the path-based algorithm over the three domain graphs (\textit{G-Baseline}, \textit{G-New}, and \textit{G-Relaxed}). We implemented the path-based inference algorithm with Apache Giraph, running on a cluster with 2 nodes (each with 48 2.7-GHz cores and 256 GB aggregated memory).

When computing true positive rate (TPR) and false positive rate (FPR), we use ten-fold cross validation. We randomly divide the malicious ground truth into ten folds and perform 10 round executions of the inference algorithm. In each round, we pickup one different fold as training set and the remaining nine folds as test set. We repeat the ten-fold testing for 5 times using different random divisions of the malicious ground truth each time, which gives a total of $50$ execution rounds of the inference algorithm. For each round, we compute the TPR and the FPR for various threshold values ($malicious_t$). For each $malicious_t$ value,  the TPR is computed as the percentage of malicious domains in the malicious test set with scores above $malicious_t$. The FPR is computed as the percentage of domains in the benign ground truth with scores above $malicious_t$. $malicious_t$ is varied between $0$ and $1$ with $0.01$ steps. We finally report TPR and FPR for each $malicious_t$ threshold value as the average over the values in the fifty rounds.

\begin{figure*}
\begin{multicols}{2}
    \begin{subfigure}{0.45\textwidth}
        \includegraphics[width=\linewidth]{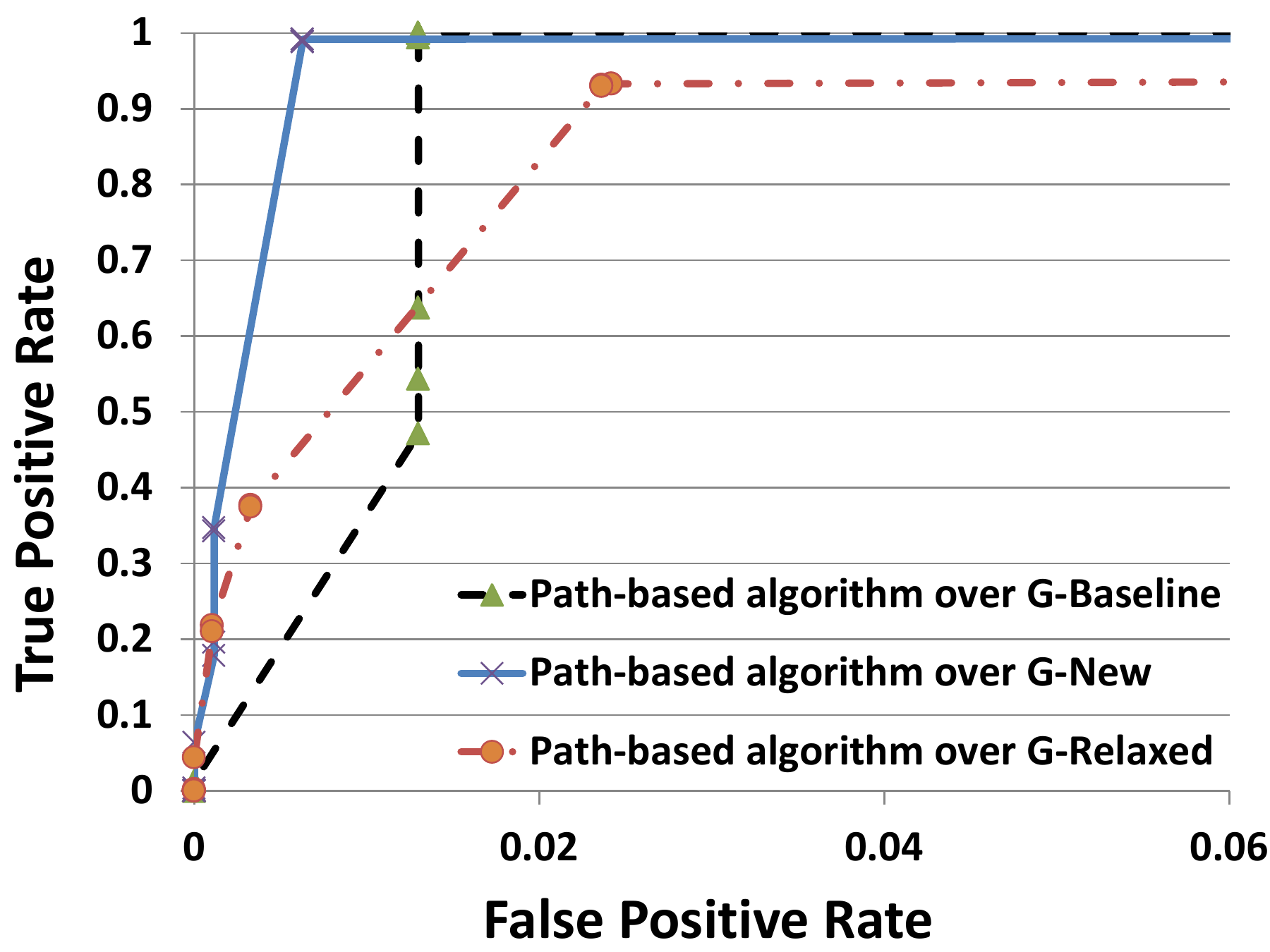} 
        \caption{dataset week-1}
        \label{fig:path_week1}
    \end{subfigure}
    \begin{subfigure}{0.45\textwidth}
        \includegraphics[width=\linewidth]{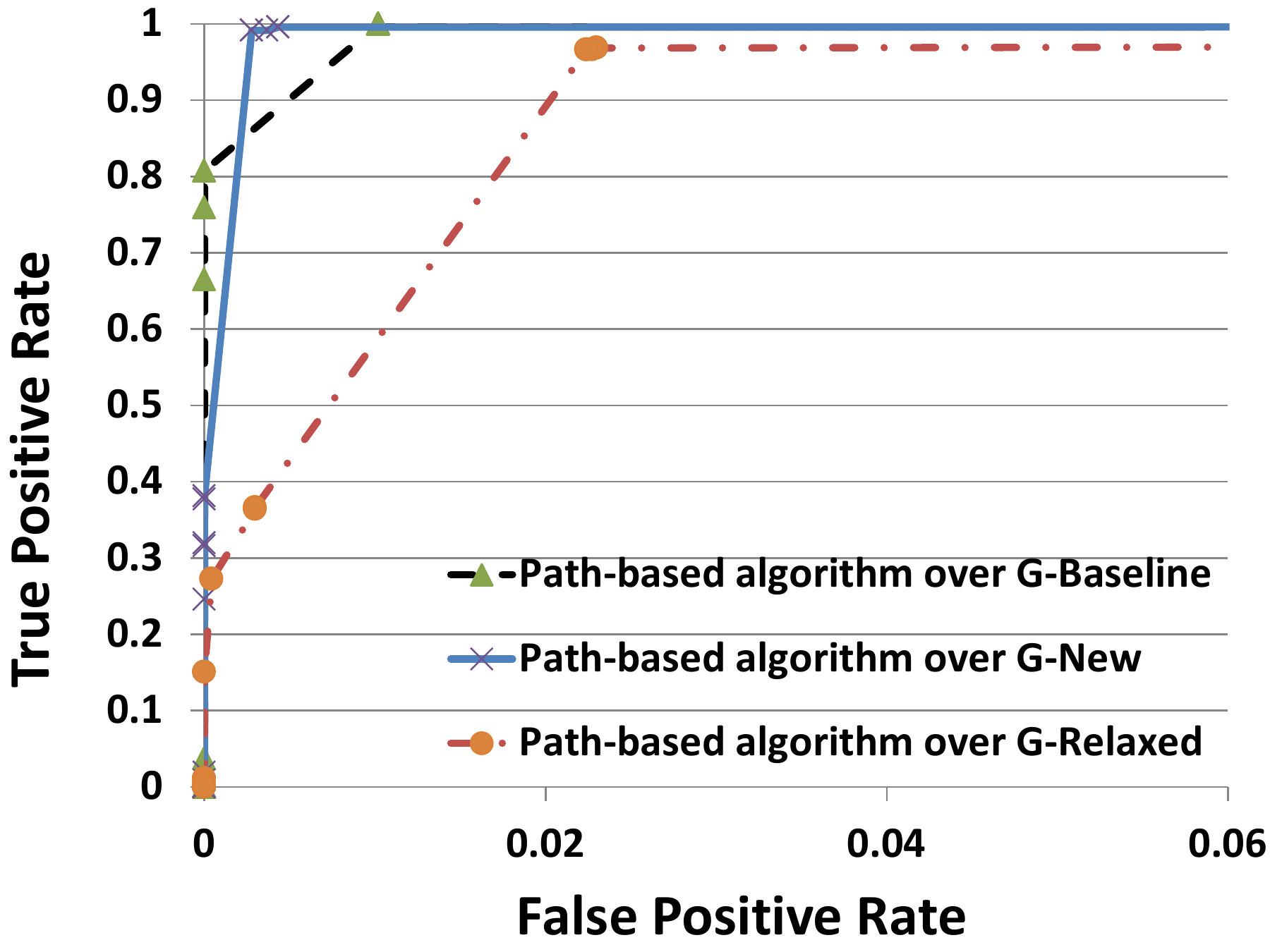} 
        \caption{dataset week-2}
        \label{fig:path_week2}
    \end{subfigure}
\end{multicols}
\vspace{-1em}
\caption{Path-based algorithm over three domain graphs}
\label{fig:path_based_graphs}
\end{figure*}

Figure~\ref{fig:path_based_graphs} shows the ROC curves of the true positive rate and the false positive rate on the three domain graphs for two different weeks. The figure clearly shows that our first type of associations results in better detection accuracy than the baseline. For example, with the same 99\% true positive rate, \textit{G-New} has a 0.63\% false positive rate compared to the 1.33\% false positive rate in \textit{G-Baseline} for the week of Feb. 4-10, 2017. Therefore, our first type of associations not only significantly expands domain coverage, but also proves to be stronger and more meaningful as evidenced by the improvement in detection accuracy. The degradation in detection accuracy in \textit{G-Baseline} is mainly due to the noise introduced by the public hosting environments. In public hosting environments, domains from unrelated owners would be hosted at the same pool of IPs. Combining this with the fact that some public hosting environments span more than one AS (e.g., both AS16509 and AS14618 belong to Amazon), would result in wrong associations. That is, two unrelated domains may still be associated even if they only use services from a single provider, which degrades the quality of the associations, and consequently affects the detection accuracy.


Figure~\ref{fig:path_based_graphs} also shows that, even though our second type of associations has the largest domain coverage (Table~\ref{tab:graph_size}), the detection accuracy on \textit{G-Relaxed} is much worse. For example, to achieve a reasonable true positive rate of 93.16\%, the false positive rate soars to 2.36\% in the active DNS data of the week of Feb. 4-10, 2017. This can be attributed to the relaxed association rule that obviously generates weak or irrelevant associations. Therefore, not all associations that expand domain coverage lead to acceptable detection accuracy. This clearly highlights the intricate challenge in designing appropriate associations that can achieve the delicate balance of increasing domain coverage while improving, or at least not considerably degrading, detection accuracy.

\begin{figure}[!t]
\centering
\includegraphics[width=3.5in]{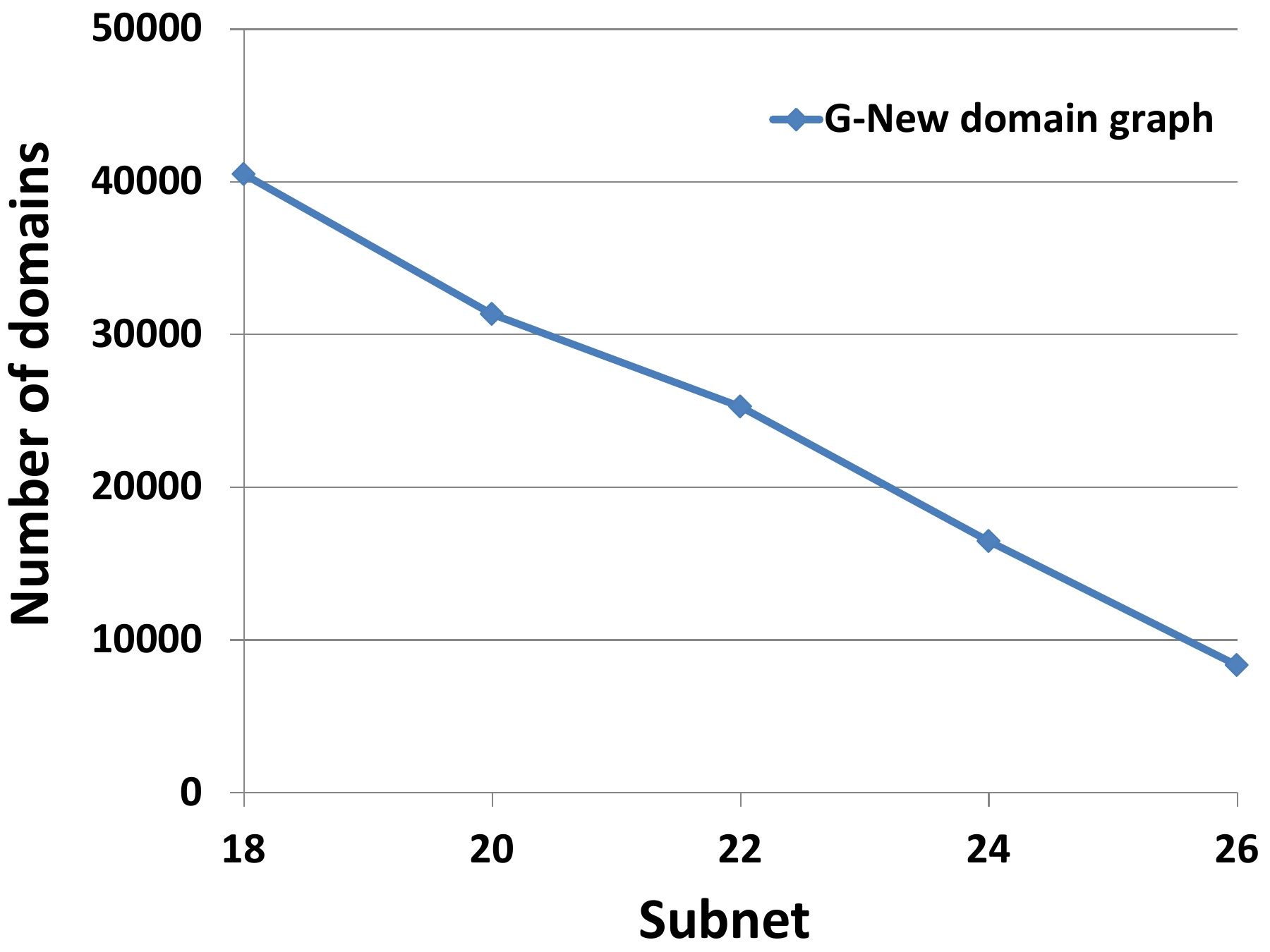}
\caption{$G-New$ graph expansion (week-1) under different subnets}
\label{fig:graph-expansion-subnet}
\end{figure}

\begin{figure}[!t]
\centering
\includegraphics[width=3.5in]{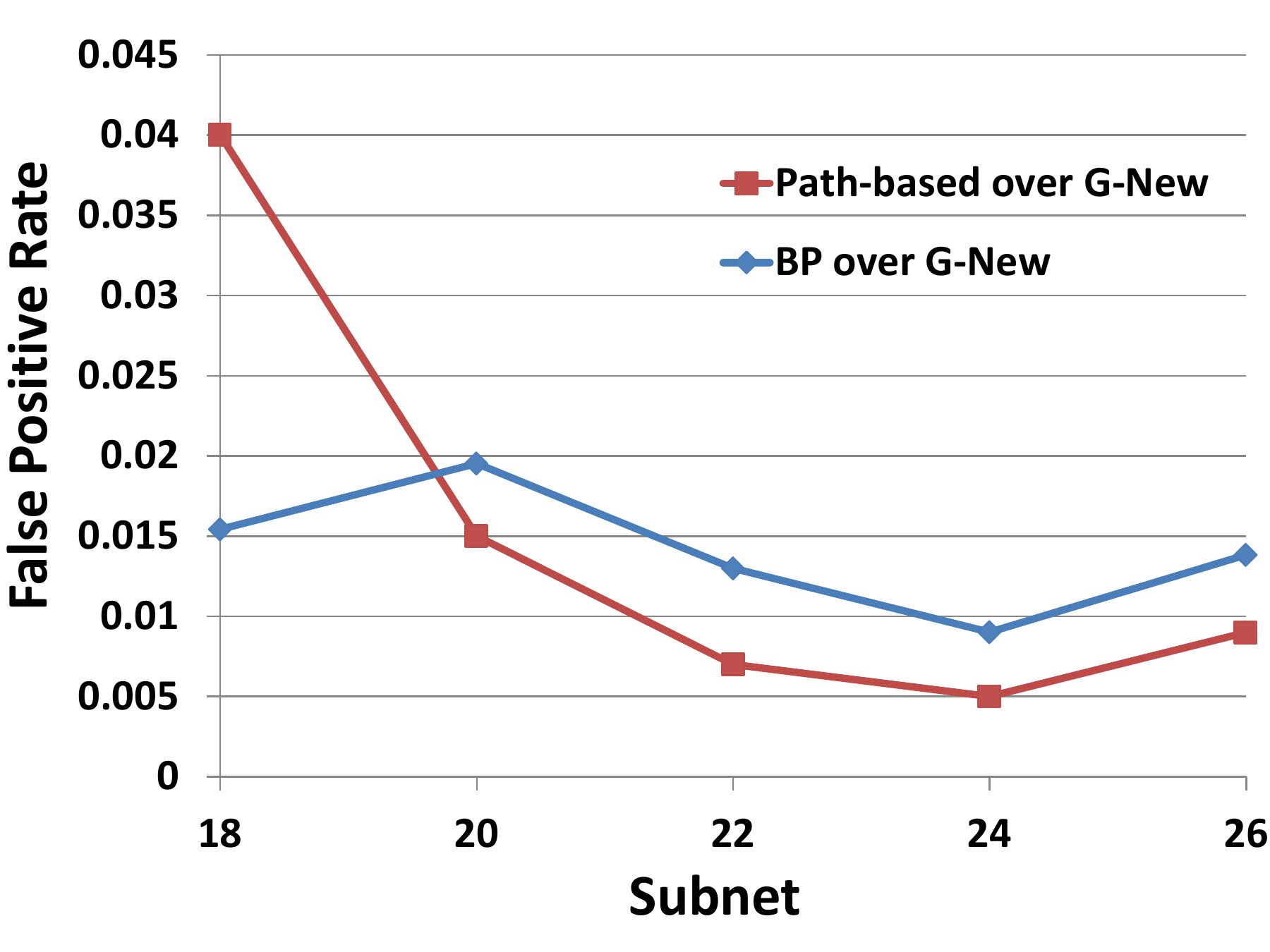}
\caption{False positive rate of the two inference algorithms on $G-New$ graph of week-1 under different subnets }
\label{fig:fpt-subnet}
\end{figure}

Before we conclude this section, we would like to provide our rationale behind using /24-subnet for the construction of domain graphs. We in fact experimented with different subnet sizes from 18 to 26 to study the effect on the expansion of the domain graph and the accuracy of malicious domain detection. As can be seen from Figure~\ref{fig:graph-expansion-subnet}, the domain coverage increases with increasing size of subnets. This is expected as with larger subnets, our association rules are able to find more domains that can be associated with one another. However, not all such associations are strong enough to accurately detect malicious domains. As shown in Figure~\ref{fig:fpt-subnet}, we get the best accuracy, i.e., the least false positive rate, when the subnet size is 24. Figure~\ref{fig:fpt-subnet} shows the false positive rate for path-based and BP algorithms on $G-New$ domain graph. One possible reason for this behavior, as mentioned earlier, is that when subnet sizes are large, it may represent IPs from multiple unrelated ASs and when subnet sizes are small, IPs for the same AS may get split into different subnets resulting in weaker or incorrect associations.

\section{Detection Accuracy and Scalability Tradeoff}\label{perf}


The path-based inference algorithm has a complexity of $O(s|V|^2)$, where $s$ is the size of the malicious seed and $V$ is the set of vertexes in \textit{G-New}. Even with the help of distributed computing platforms, it could still be quite expensive to handle large-scale DNS data. In this section, we investigate techniques to strike a good balance between detection accuracy and efficiency.

One natural alternative is BP. We implemented the BP algorithm in C program and ran it in a single multi-core server with 48 2.7-GHz cores and 256 GB memory. Our experiments follow the convergence rules of BP that are mentioned earlier in Section~\ref{background}  with the convergence threshold and the maximum number of iterations are empirically selected as $1\times 10^{-10}$ and 15, respectively.

\begin{figure}[!t]
\centering
\includegraphics[width=3.5in]{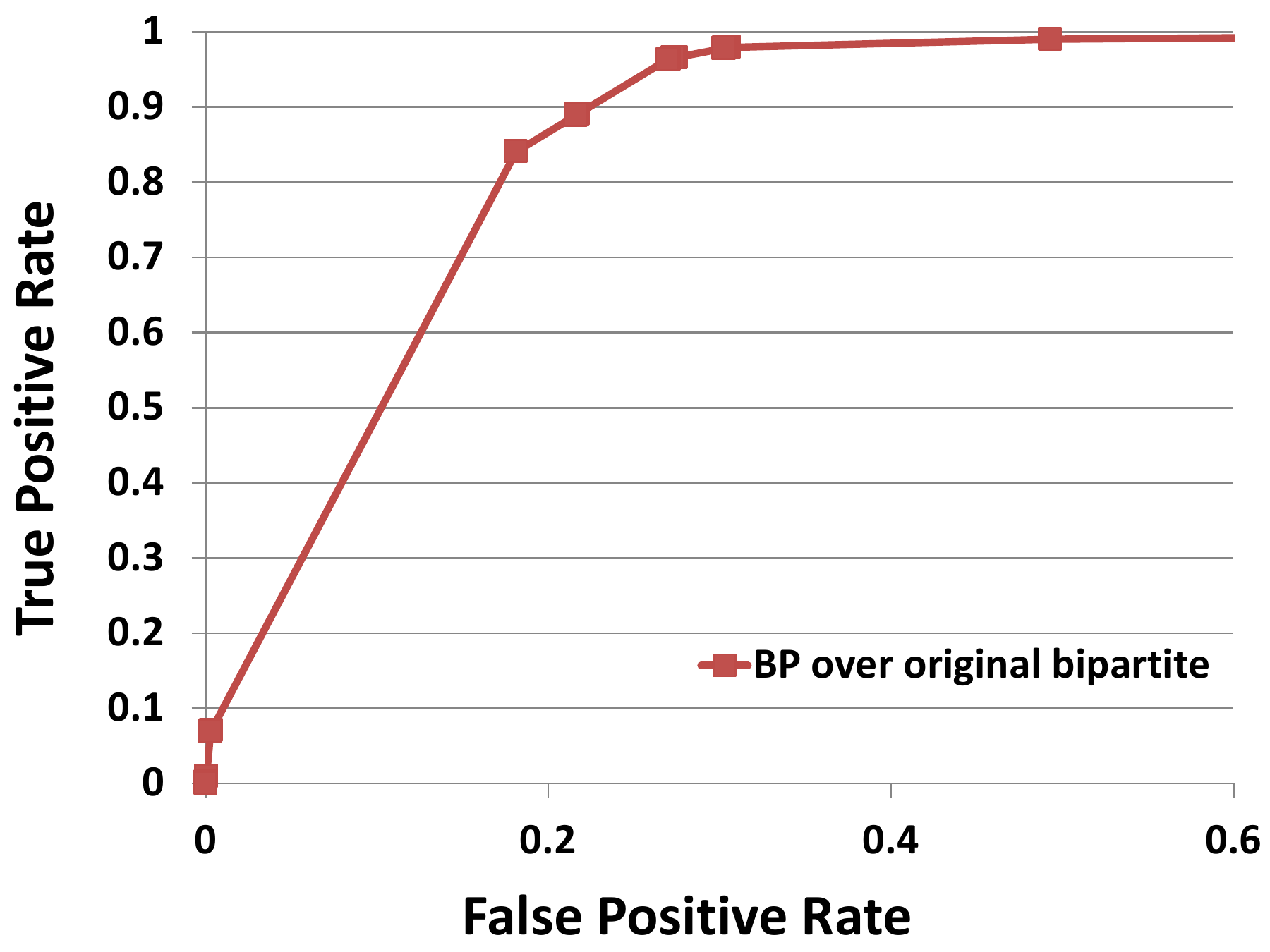}
\caption{BP on bipartite graph of whole DNS data of week-1}
\label{fig:bp-original-bipartite}
\end{figure}

As shown in Figure~\ref{fig:bp-original-bipartite}, our experiments confirm what has been shown before~\cite{GuiltyByAssociation_Khalil2016} that applying BP directly on the bipartite graph corresponding to the whole DNS data yields very poor detection accuracy. For example, applying BP on week-1 results in 49\% false positive rate for a true positive rate of 99\%. The reason is that we cannot conclude the maliciousness of an IP simply because a malicious domain is resolved to it. In other words, hosting relationships alone are not strong enough to reliably reason the maliciousness of unknown domains and IPs. As shown in Section~\ref{construction}, the enhanced domain graph captures much stronger relationships between domains. It would be compelling to investigate how BP on domain graphs could assist in producing results with acceptable accuracy with much less computational cost.

As in each round of BP a single message is passed along each edge, the complexity of one-round BP is simply $O(|E|)$, where $E$ is the set of edges. In a sparse graph, $|E|$ is multi-magnitude smaller than $|V|^2$. Though in the worst case, many rounds of propagation has to be performed until convergence\footnote{In fact for some special graphs, BP may never converge}, in practice it is often sufficient to stop after a pre-determined constant number of rounds (e.g., 15 or 20). Thus, BP over the domain graph could be much more efficient than using the path-based algorithm. 


Note that the path-based algorithm is specifically designed for malicious domain detection over domain graphs (e.g., the explicit decay mechanism and the particular way of combining inferences from multiple malicious seeds). BP on the other hand is a generic inference algorithm that could only implicitly reflect some of the intuitions behind the path-based approach (e.g., the influence of a node is diminishing when a message is passed along a long path), which may lead to lower detection accuracy. 

We also explore another possible way to apply BP over \textit{G-New}, based on the following observation. Suppose that an entity hosts $n$ domains on its server with a dedicated IP. Reflected in \textit{G-New}, we will have a clique among these $n$ domains, with $n(n-1)/2$ edges. Meanwhile, in the original bipartite domain resolution graph, there will be only $n$ edges between these domains and the dedicated IP. In general, \textit{G-New} would be much denser than the bipartite domain resolution graph. Since the complexity of BP is proportional to the number of edges, to further improve efficiency, another alternative is to run BP on the bipartite graph reduced by the domains in \textit{G-New}. In detail, given the original domain resolution graph, we keep an edge between a domain $d$ and an IP only if $d$ is in \textit{G-New}. The resulting bipartite graph has the same set of domains as \textit{G-New}, but would be much sparser. Therefore, running BP over this reduced bipartite graph would also be more efficient than over \textit{G-New}. 

This approach, however, could cause further deterioration of detection accuracy, due to several reasons. First, inferences become less direct. An edge in \textit{G-New} is now corresponding to multiple indirect paths due to intermediate IP nodes in the bipartite graph. The longer the paths, the weaker the inference. Second, probably more importantly, the induced bipartite graph could introduce unreliable associations. For instance, consider two domains $d_1$ and $d_2$ who only share a single public IP $i$, though they may share dedicated IPs or IPs from different ASs with other domains, and thus appear in \textit{G-New}. In \textit{G-New} there would be no direct edge between them. However, in the induced graph, we will have edges $(d_1, i)$ and $(d_2, i)$, which could cause unwanted inference when applying BP.

\begin{figure*}
\begin{multicols}{2}
    \begin{subfigure}{0.45\textwidth}
        \includegraphics[width=\linewidth]{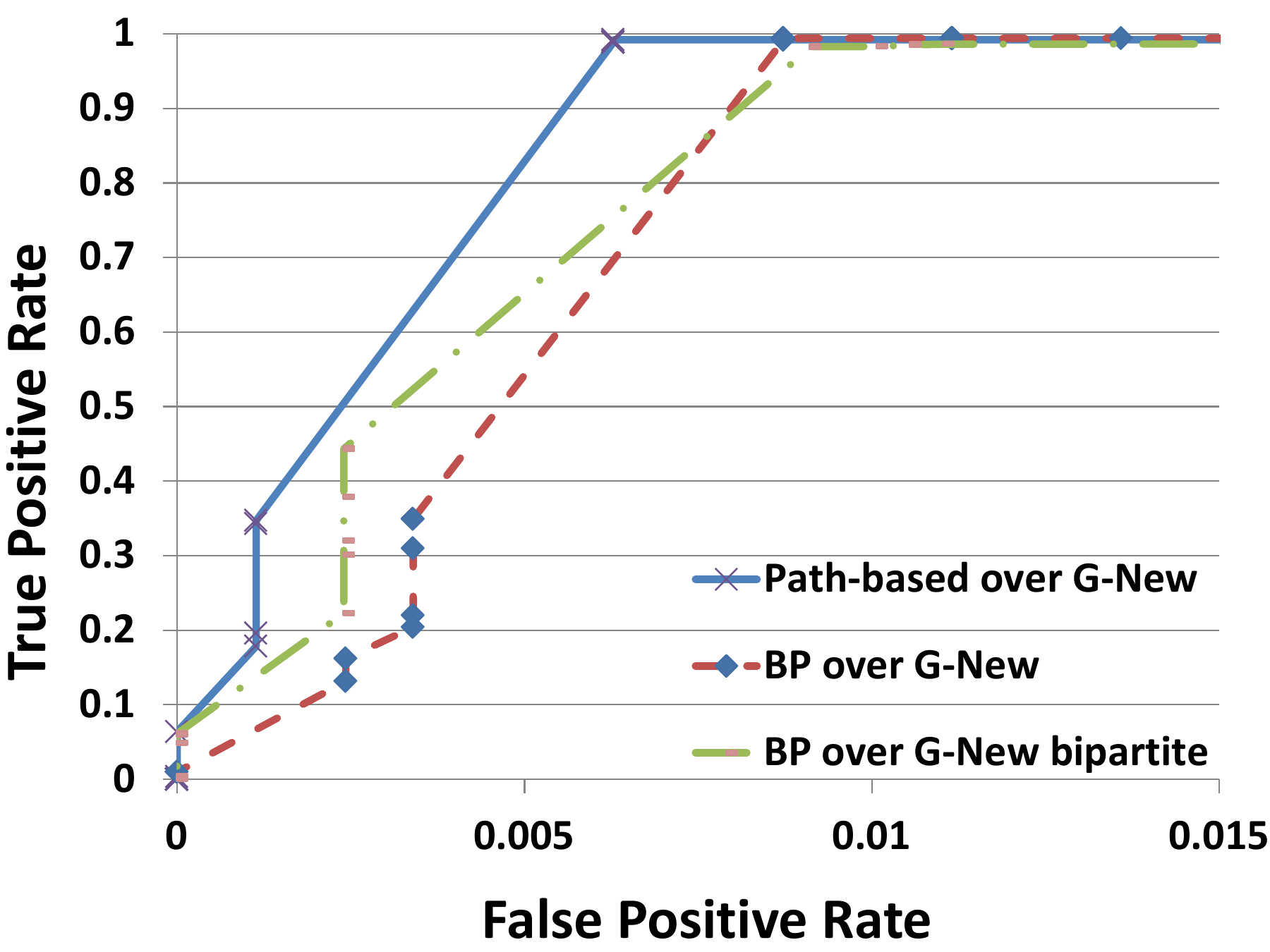} 
        \caption{dataset week-1}
        \label{fig:compare_week1}
    \end{subfigure}
    \begin{subfigure}{0.45\textwidth}
        \includegraphics[width=\linewidth]{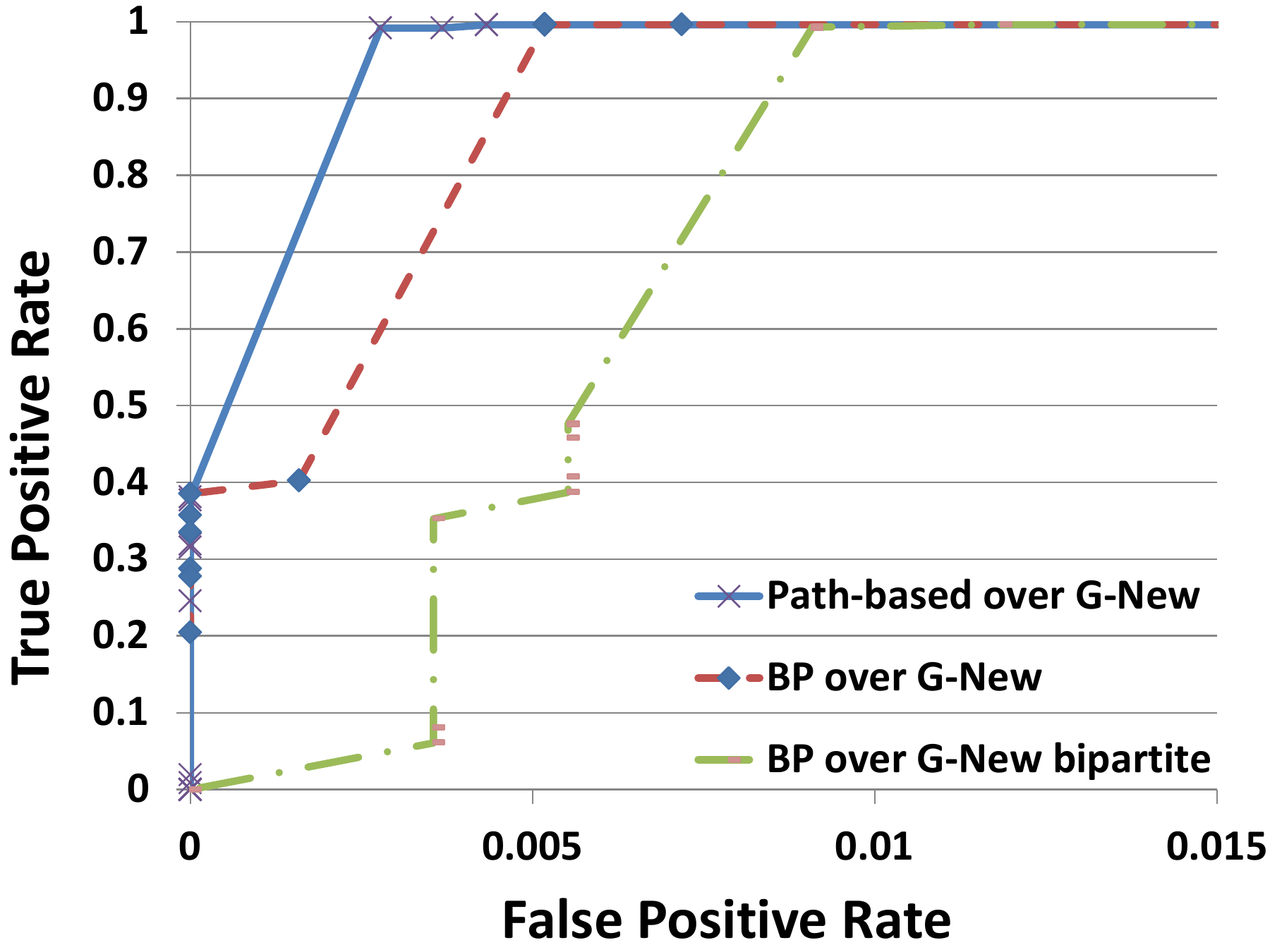} 
        \caption{dataset week-2}
        \label{fig:compare_week2}
    \end{subfigure}
\end{multicols}
\caption{Detection accuracy of Path-based algorithm and BP algorithm}
\label{fig:perf-detection}
\vspace{-2em}
\end{figure*}

Figure~\ref{fig:perf-detection} shows ROC curves of the three approaches: the path-based approach and BP over \textit{G-New}, and BP over the induced bipartite graph. 
Our first observation is that all three approaches achieve high detection accuracy. For both datasets, they achieve more than $99\%$ true positive rates with less than $1\%$ false positive rates. This demonstrates the effectiveness of our proposed association scheme. Since it captures accurately the connection between domains, even with generic inference techniques, malicious domains could still be identified accurately. Second, the path-based approach indeed offers superior detection accuracy in both datasets, which is consistent with our intuition. Depending on the datasets, with the same high true positive rate (over $99\%$), the path-based inference algorithm's false positive rates are one-third or half of that of the approaches using BP. For the two approaches using BP, in general, the one running over \textit{G-New} offers a little better accuracy, though it could be insignificant, depending on the datasets.

\begin{figure}[!t]
\centering
\includegraphics[width=3.5in]{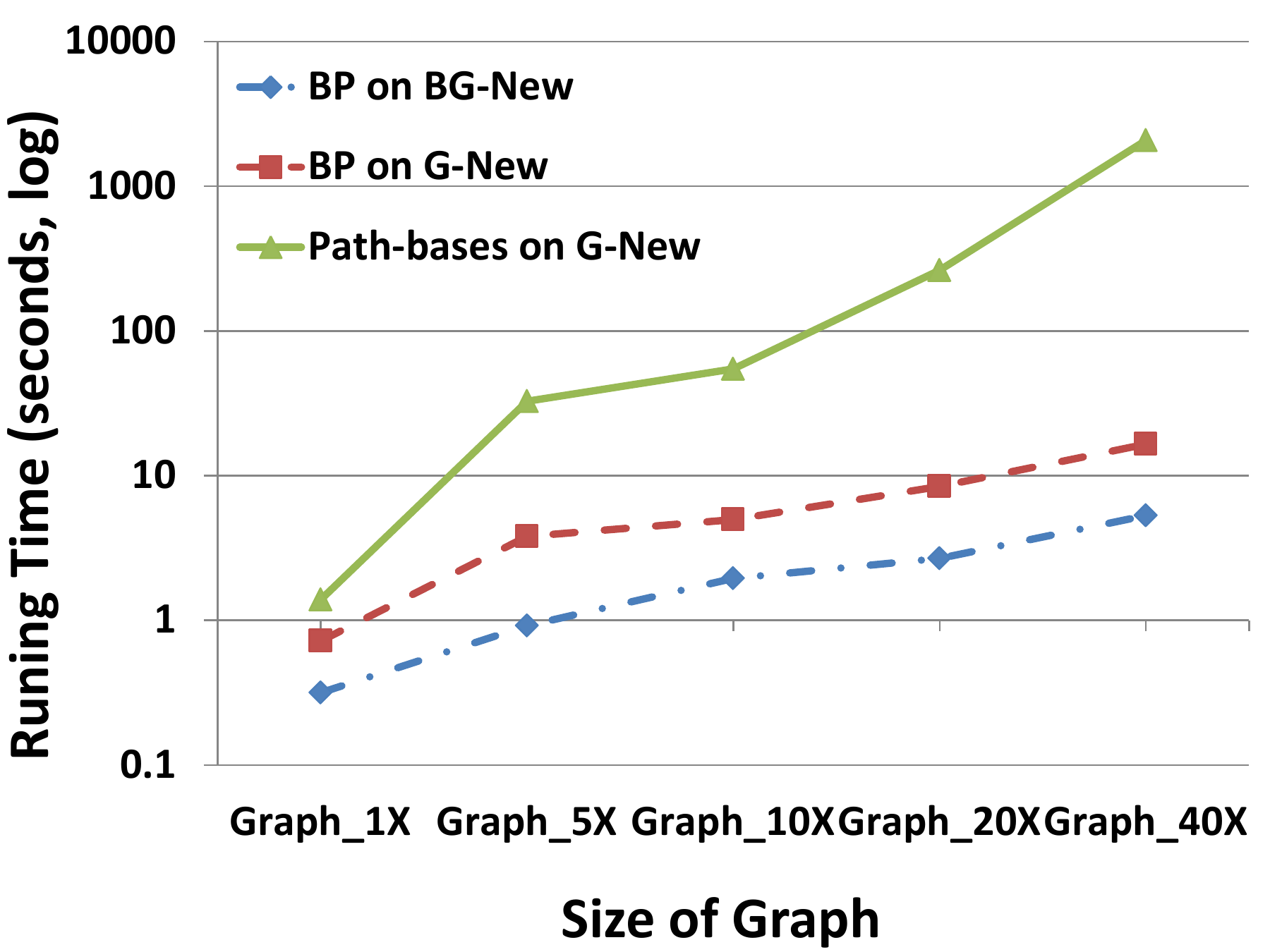}
\caption{Running time of Path-based and BP algorithms}
\label{fig:perf-scale}
\vspace{-2em}
\end{figure}

Given the seemingly rather small advantage of the path-based approach in terms of false positive rates, one may wonder whether it matters in practice. We note that, due to base-rate fallacy~\cite{Axelsson00}, it could indeed have a significant impact. For example, suppose 98\% of all checked domains are benign (the base rate), with the same high true positive rate (99\%), a 0.2\% false positive rate would result in 90\% precision (i.e., the percentage of the detected domains that are indeed malicious). A seemingly slight increase of the false positive rate to 0.6\% would reduce the precision to 70\%.

Next, we compare their efficiency and scalability when handling large-scale DNS data. For this purpose, we generate synthetic domain resolution graphs with different scales (5$\times$, 10$\times$, 20$\times$ and 40$\times$ respectively) based on a real domain resolution graph (Feb. 20-26, 2017). The graph generation algorithm ensures that the synthetic graphs have the same degree distribution as the real one.

Figure~\ref{fig:perf-scale} shows the running time (in log scale) of the three approaches when we scale up the DNS dataset. Note that the running time reported here excludes the time of other overheads (e.g., loading graphs to the distributed file system in Giraph) to clearly illustrate the difference of their computational costs. We see that though all three algorithms can handle the small graph (Graph\_1$\times$) efficiently, the running time of the path-based algorithm increases rapidly when we scale up the graph. For example, when handling the domain graph (with 751,160 nodes and 4,129,156 edges) derived from Graph\_40$\times$, it takes the path-based approach more than 2000 seconds to finish, while the running times of the two BP approaches are below 20 seconds. BP over the induced bipartite graph (\textit{BG-New}) clearly offers the best performance due to the reduced number of edges. 

Figures~\ref{fig:perf-detection} and~\ref{fig:perf-scale} suggest a clear tradeoff between the three approaches: with sufficient computational resources and when handling moderate scale data, the path-based approach would yield the best detection accuracy. When handling large-scale data, applying BP over \textit{G-New} or the induced bipartite graphs would offer practical scalable solutions with good detection accuracy.

\section{Related Work}\label{related}
In this section, we compare and contrast our work with the previous research in the same area.

\textbf{IP Address Classification:} 
Historically, IP address classification is motivated by the need to efficiently allocate IPs to different organizations as well as to improve the efficiency of routing IP packets from one router to another~\cite{Ruiz-Sanchez:2001:IPLookup}. Xie et al.~\cite{Xie:2007:DynamicIP} propose a technique to automatically classify IP addresses as dynamic, that is, DHCP allocated IPs, and static using Hotmail server logs in order to identify spams. Their observation is that most of the spam email servers are hosted at dynamic IP addresses. However, this observation is not reflected on the malicious domains we identified and, thus, such an IP classification does not assist us in constructing domain graphs with strong associations. Recently, Scott et al.~\cite{Scott:2016:CDNIP} propose to capture the IP footprint of CDN (content delivery network) deployments. While this work identifies IPs of shared infrastructures, these IPs may not necessarily be public IPs as we classify in our work in order to detect malicious domains. IP ranges of  Cloud computing platforms, such as Amazon AWS~\cite{AWSPublicIP}, Microsoft Azure~\cite{AzurePublicIPList} and Google Cloud~\cite{GooglePublicIP}, are usually available for everyone. However, not all these IPs are public as we define in this study. To the best of our knowledge, we are the first to accurately classify IPs in the wild as public and dedicated.

\textbf{Malicious Domain Detection:}
There is a vast body of research devoted to detecting malicious domains via static analysis. Such research work can be classified into two techniques, host-based and network-based. Briefly, host-based approaches rely on detecting malware signatures in programs running on end hosts~\cite{Kolbitsch:2009:EEM,Rieck:2011:AAM}, whereas network-based approaches rely on detecting specific patterns and fingerprints by monitoring the network traffic~\cite{Antonakakis:2011:DMD,GuiltyByAssociation_Khalil2016}. Since our approach is network-based, we compare and contrast the most relevant network-based proposals which rely on DNS data for malicious domain detection. Network-based approaches can further be divided into classification based~(e.g., \cite{DetectingMaliciousActivityWithDnsBackscatter_Fukuda2015,Segugio_Rahbarinia2015,DeepDGA_Anderson2016,MethodForDetectingDgaBotnet_Tong2016,Notos_Antonakakis2010,Exposure_Bilge2014}) and inference based approaches~(e.g., ~\cite{GuiltyByAssociation_Khalil2016,Smash_Zhang2015,DetectingMaliciousDomainsViaGraphInference_Manadhata2014,Tamersoy:2014:GAL}). While the classification based approaches primarily rely on local network and host information, inference based approaches exploit the global relationships among domains along with local information in order to better detect malicious domains. Our work falls in the latter category. 
Now we compare and contrast our work with respect to these two categories below.

\textbf{Classification based approaches.} Many approaches~\cite{DetectingMaliciousActivityWithDnsBackscatter_Fukuda2015,Segugio_Rahbarinia2015,DeepDGA_Anderson2016,MethodForDetectingDgaBotnet_Tong2016}, including Notos~\cite{Notos_Antonakakis2010} and EXPOSURE~\cite{Exposure_Bilge2014}, identify malicious domains by building a classifier using the local features extracted from passive DNS data along with other network information such as WHOIS records~\cite{Liu:2015:CLP}. Such approaches are effective as long as the local features used in the classification are not manipulated. However, it has been shown~\cite{TowardsSystematicEvaluationOfTheEvadabilityOfBotnetDetectionMethods_Stinson2008} that many local features such as TTL based features and patterns in domain names, are easy to manipulate and thus rendering such techniques less effective. These approaches perform best when one has access to sensitive individual DNS queries which are difficult to gain access to. On the other hand, inference based approaches like ours can detect malicious domains with high accuracy using only aggregate DNS data which is relatively easier to gain access to.

\textbf{Inference based approaches.} 
Inference based approaches have been proposed to complement classification based approaches by considering not only local network features but also the associations among domains. We already discussed the related work by Manadhata et al.~\cite{DetectingMaliciousDomainsViaGraphInference_Manadhata2014} and Khalil et al.~\cite{GuiltyByAssociation_Khalil2016} in Section~\ref{introduction} and Section~\ref{background}.

Zou et al.~\cite{DetectingMalwareBasedOnDnsGraphMining_Zou2015} proposed a similar approach based on BP but they utilized domain-IP associations in addition to domain-host associations in order to build the graph. Active DNS data used in our study can also be modeled as a bipartite graph and then BP can be applied the bipartite graph. However, we observe that the accuracy of the inference is unacceptably low as the associations in the Active DNS data are much more weaker than those in DNS query logs.

SMASH~\cite{Smash_Zhang2015} is an unsupervised approach to infer groups of related servers involved in malware campaigns.  It focuses on server side communication patterns extracted from HTTP traffic to systematically mine relations among servers from multiple dimensions. SMASH is novel in proposing a mechanism that utilizes connections among malicious severs to detect malware campaigns in contrast to classification schemes that solely use individual server features. Our approach is similar to SMASH in establishing server associations as bases for identifying new malicious servers, but complements SMASH by utilizing active DNS data, in contrast to HTTP traffic, which offers privacy benefits as  active DNS data is publicly available database and has no privacy or security liability associated with it. Additionally, instead of using second-level domain names, our approach establishes associations among fully qualified domain names. This relaxes the assumption in SMASH that servers with the same second-level domain belong to the same organization and hence, our approach detects malicious dynamic DNS servers.

Additionally, Gao et al.~\cite{ReexaminingDnsFromAGlobalRecursiveResolverPerspective_Gao2016} propose an approach to detect malicious domains exploiting temporal correlations in DNS queries. Rahbarinia et al.~\cite{Segugio_Rahbarinia2015} constructs a host-to-domain bipartite graph to efficiently detect new malicious domains by tracking the DNS query behavior using DNS data collected from within large scale ISP networks. 
Stevanovic et al.~\cite{MethodForIdentifyingCompromisedClients_Stevanovic2017} identify compromised hosts by analyzing DNS traffic traces from different operational ISP networks. All of these approaches rely on individual DNS queries made by users which are not readily available with aggregate data sources such as Active DNS datasets we use in this study.

Very recently, Alrwais et al.~\cite{Alrwais:2017:MD} analyze and detect bulletproof hosting (BPH) services, which provide Internet miscreants with infrastructure that is resilient to complaints of illicit activities, on legitimate hosting providers. They shed lights on how BPH services have moved from self-managed monolithic infrastructure to the sub-allocations within third-party hosting services in order to evade reputation based detection such as BGP ranking and ASwatch~\cite{Konte:2015:ARS}. They detect malicious sub-allocations within ASs as opposed to malicious ASs~\cite{Chen_2008_MAS,Stone-Gross:2009:FFR}. They rely on two key datasets: Whois dataset that is used to spot sub-allocations, and Passive
DNS dataset, that is used to extract signals indicating malicious behavior. While their approach detects a very specific subset of malicious IPs, our approach is designed to detect any malicious domains in the wild that behave similar to known malicious domains. We believe that their detection accuracy could be improved by adopting our techniques of detecting malicious domains.

\textbf{Dynamic Malware Analysis:} Unlike the static malware analysis discussed above, dynamic malware analysis usually consists of executing sample malware artifacts in a secure, controlled and isolated environment. Such executions provide a complementary means of studying the behavior of malware and how it interact with the system including the network and file system. A number of studies~\cite{Rossow:2012:LAM,Wang:2014:WPM,Nappa:2013:DCA} have analyzed rouge infrastructure used to distribute malware in order to detect malicious systems including command and control servers and malware distribution servers. More recently, Lever et al.~\cite{Lever:2017:PDNS} analyze using several large malware and network datasets that span across 5 years. They show that Internet miscreants are increasingly using potentially unwanted programs (PUPs)~\cite{Kotzias:2015:CPA,Thomas:2015:AIS} to launch attacks and build classifiers to differentiate between PUPs from malware domains. As previous research has confirmed, they show at a large scale that, for the majority of malware samples they studied, network traffic provides the earliest indicator of infection which is in fact several weeks and even often months before the malware sample is discovered. This shows that frequent execution of reputation based approaches using DNS data like ours could identify and build up-to-date blacklists helping in the cause of blocking and mitigating malware.

\section{Conclusion}\label{conclusion}
Active DNS data is an important and easily accessible information source for malicious domain detection. However, compared to other types of DNS data, active DNS data offer limited information. Overcoming the inherent limitations, we design a novel domain association scheme based on active DNS data to facilitate inference-based malicious domain detection. Our scheme relies on a deep analysis of dedicated and public IPs, which significantly improves over existing domain association schemes in terms of not only domain coverage but also detection accuracy. We show that our scheme could be integrated with both specific path-based inference algorithms and generic inference algorithms such as BP. We further explore ways to improve the efficiency and scalability of malicious domain detection, and carefully study the tradeoff with detection accuracy. For future work, it would be interesting to investigate means to integrate active DNS data with other pubic data sources (e.g, domain authoritative servers and WHOIS records) and further strengthen the quality of domain associations. Another possible avenue is to empirically evaluate the effectiveness of our association scheme with other types of DNS data.

\end{document}